\documentclass[aps,prd,twocolumn,superscriptaddress,preprintnumbers,showpacs,amsmath,amssymb,nofootinbib]{revtex4}

\usepackage{graphicx}
\usepackage{dcolumn}
\usepackage{bm}
\usepackage[dvips]{color}
\usepackage{epsfig}
\usepackage{xspace}

\begin{document}

\newcommand{\mbc}{\ensuremath{M_{\rm bc}}\xspace}
\newcommand{\de}{\ensuremath{\Delta E}\xspace}
\newcommand{\thr}{\ensuremath{\cos\theta_{\rm thr}}\xspace}
\newcommand{\fish}{\ensuremath{\mathcal{F}}\xspace}

\newcommand{\bdk}{\ensuremath{B^{\pm}\to DK^{\pm}}\xspace}
\newcommand{\bdkm}{\ensuremath{B^{-}\to DK^{-}}\xspace}
\newcommand{\bdkp}{\ensuremath{B^{+}\to DK^{+}}\xspace}
\newcommand{\bdtk}{\ensuremath{B^{\pm}\to \tilde{D}K^{\pm}}\xspace}
\newcommand{\bdtkp}{\ensuremath{B^{+}\to \tilde{D}_+K^{+}}\xspace}
\newcommand{\bdtkm}{\ensuremath{B^{-}\to \tilde{D}_-K^{-}}\xspace}

\newcommand{\bdsk}{\ensuremath{B^{\pm}\to D^{*}K^{\pm}}\xspace}
\newcommand{\bdskm}{\ensuremath{B^{-}\to D^{*}K^{-}}\xspace}
\newcommand{\bdskp}{\ensuremath{B^{+}\to D^{*}K^{+}}\xspace}
\newcommand{\bdstk}{\ensuremath{B^{\pm}\to \tilde{D}^{*}K^{\pm}}\xspace}
\newcommand{\bdstkp}{\ensuremath{B^{+}\to \tilde{D}^{*}_+K^{+}}\xspace}
\newcommand{\bdstkm}{\ensuremath{B^{-}\to \tilde{D}^{*}_-K^{-}}\xspace}

\newcommand{\bdks}{\ensuremath{B^{\pm}\to DK^{*\pm}}\xspace}
\newcommand{\bdksm}{\ensuremath{B^{-}\to DK^{*-}}\xspace}
\newcommand{\bdksp}{\ensuremath{B^{+}\to DK^{*+}}\xspace}
\newcommand{\bdtks}{\ensuremath{B^{\pm}\to \tilde{D}K^{*\pm}}\xspace}
\newcommand{\bdtksp}{\ensuremath{B^{+}\to \tilde{D}_+K^{*+}}\xspace}
\newcommand{\bdtksm}{\ensuremath{B^{-}\to \tilde{D}_-K^{*-}}\xspace}

\newcommand{\bddsk}{\ensuremath{B^{\pm}\to D^{(*)}K^{\pm}}\xspace}
\newcommand{\bddskm}{\ensuremath{B^{-}\to D^{(*)}K^{-}}\xspace}
\newcommand{\bddskp}{\ensuremath{B^{+}\to D^{(*)}K^{+}}\xspace}
\newcommand{\bddstk}{\ensuremath{B^{\pm}\to \tilde{D}^{(*)}K^{\pm}}\xspace}
\newcommand{\bddstkp}{\ensuremath{B^{+}\to \tilde{D}^{(*)}_+K^{+}}\xspace}
\newcommand{\bddstkm}{\ensuremath{B^{-}\to \tilde{D}^{(*)}_-K^{-}}\xspace}

\newcommand{\bddsks}{\ensuremath{B^{\pm}\to D^{(*)}K^{(*)\pm}}\xspace}
\newcommand{\bddsksm}{\ensuremath{B^{-}\to D^{(*)}K^{(*)-}}\xspace}
\newcommand{\bddsksp}{\ensuremath{B^{+}\to D^{(*)}K^{(*)+}}\xspace}
\newcommand{\bddstks}{\ensuremath{B^{\pm}\to \tilde{D}^{(*)}K^{(*)\pm}}\xspace}
\newcommand{\bddstksp}{\ensuremath{B^{+}\to \tilde{D}^{(*)}_+K^{(*)+}}\xspace}
\newcommand{\bddstksm}{\ensuremath{B^{-}\to \tilde{D}^{(*)}_-K^{(*)-}}\xspace}

\newcommand{\bdksnr}{\ensuremath{B^{\pm}\to DK^0_S\pi^{\pm}}\xspace}
\newcommand{\bdkspnr}{\ensuremath{B^{+}\to DK^0_S\pi^{+}}\xspace}
\newcommand{\bdksmnr}{\ensuremath{B^{-}\to DK^0_S\pi^{-}}\xspace}
\newcommand{\bdtksnr}{\ensuremath{B^{\pm}\to \tilde{D}K^0_S\pi^{\pm}}\xspace}
\newcommand{\bdtkspnr}{\ensuremath{B^{+}\to \tilde{D}_+K^0_S\pi^{+}}\xspace}
\newcommand{\bdtksmnr}{\ensuremath{B^{-}\to \tilde{D}_-K^0_S\pi^{-}}\xspace}

\newcommand{\bdpi}{\ensuremath{B^{\pm}\to D\pi^{\pm}}\xspace}
\newcommand{\bdpip}{\ensuremath{B^{+}\to D\pi^{+}}\xspace}
\newcommand{\bdpim}{\ensuremath{B^{-}\to D\pi^{-}}\xspace}
\newcommand{\bdtpi}{\ensuremath{B^{\pm}\to \tilde{D}\pi^{\pm}}\xspace}
\newcommand{\bdtpip}{\ensuremath{B^{+}\to \tilde{D}_+\pi^{+}}\xspace}
\newcommand{\bdtpim}{\ensuremath{B^{-}\to \tilde{D}_-\pi^{-}}\xspace}

\newcommand{\bdstpi}{\ensuremath{B^{\pm}\to \tilde{D}^{*}\pi^{\pm}}\xspace}
\newcommand{\bdstpip}{\ensuremath{B^{+}\to \tilde{D}^{*}_+\pi^{+}}\xspace}
\newcommand{\bdstpim}{\ensuremath{B^{-}\to \tilde{D}^{*}_-\pi^{-}}\xspace}
\newcommand{\bdspi}{\ensuremath{B^{\pm}\to D^{*}\pi^{\pm}}\xspace}
\newcommand{\bdspim}{\ensuremath{B^{-}\to D^{*}\pi^{-}}\xspace}
\newcommand{\bdspip}{\ensuremath{B^{+}\to D^{*}\pi^{+}}\xspace}

\newcommand{\bndspi}{\ensuremath{B\to D^{*\pm}\pi^{\mp}}\xspace}
\newcommand{\bndspim}{\ensuremath{B\to D^{*+}\pi^{-}}\xspace}
\newcommand{\bndspip}{\ensuremath{B\to D^{*-}\pi^{+}}\xspace}

\newcommand{\bddspi}{\ensuremath{B^{\pm}\to D^{(*)}\pi^{\pm}}\xspace}
\newcommand{\bddspip}{\ensuremath{B^{+}\to D^{(*)}\pi^{+}}\xspace}
\newcommand{\bddspik}{\ensuremath{B^{-}\to D^{(*)}\pi^{-}}\xspace}
\newcommand{\bddstpi}{\ensuremath{B^{\pm}\to \tilde{D}^{(*)}\pi^{\pm}}\xspace}
\newcommand{\bddstpip}{\ensuremath{B^{+}\to \tilde{D}^{(*)}_+\pi^{+}}\xspace}
\newcommand{\bddstpim}{\ensuremath{B^{-}\to \tilde{D}^{(*)}_-\pi^{-}}\xspace}

\newcommand{\dsdpi}{\ensuremath{D^{*\pm}\to D\pi^{\pm}}\xspace}
\newcommand{\dsdpis}{\ensuremath{D^{*\pm}\to D\pi_s^{\pm}}\xspace}
\newcommand{\dsdpip}{\ensuremath{D^{*+}\to D^0\pi^{+}}\xspace}
\newcommand{\dsdpips}{\ensuremath{D^{*+}\to D^0\pi_s^{+}}\xspace}
\newcommand{\dsdpim}{\ensuremath{D^{*-}\to \overline{D}{}^0\pi^{-}}\xspace}
\newcommand{\dsdpims}{\ensuremath{D^{*-}\to \overline{D}{}^0\pi_s^{-}}\xspace}

\newcommand{\dsndpi}{\ensuremath{D^{*}\to D\pi^0}\xspace}
\newcommand{\dsndg}{\ensuremath{D^{*}\to D\gamma}\xspace}

\newcommand{\kspp}{\ensuremath{K^0_S\pi^+\pi^-}\xspace}
\newcommand{\dkpp}{\ensuremath{\overline{D}{}^0\to K^0_S\pi^+\pi^-}\xspace}
\newcommand{\dtkpp}{\ensuremath{D\to K^0_S\pi^+\pi^-}\xspace}

\newcommand{\dn}{\ensuremath{D^0}\xspace}
\newcommand{\db}{\ensuremath{\overline{D}{}^0}\xspace}

\newcommand{\bmu}{\boldsymbol{\mu}}
\newcommand{\bz}{\boldsymbol{z}}

\renewcommand{\deg}{\ensuremath{{}^{\circ}}\xspace}

\preprint{Belle preprint 2010-4}

\title{Evidence for direct \boldmath{$CP$} violation in the decay
\boldmath{$B^{\pm}\to D^{(*)}K^{\pm}$}, \boldmath{$D\to K^0_S\pi^+\pi^-$} and measurement of the CKM phase $\phi_3$}

\affiliation{Budker Institute of Nuclear Physics, Novosibirsk}
\affiliation{Faculty of Mathematics and Physics, Charles University, Prague}
\affiliation{University of Cincinnati, Cincinnati, Ohio 45221}
\affiliation{Department of Physics, Fu Jen Catholic University, Taipei}
\affiliation{Justus-Liebig-Universit\"at Gie\ss{}en, Gie\ss{}en}
\affiliation{The Graduate University for Advanced Studies, Hayama}
\affiliation{Hanyang University, Seoul}
\affiliation{University of Hawaii, Honolulu, Hawaii 96822}
\affiliation{High Energy Accelerator Research Organization (KEK), Tsukuba}
\affiliation{Institute of High Energy Physics, Chinese Academy of Sciences, Beijing}
\affiliation{Institute of High Energy Physics, Vienna}
\affiliation{Institute of High Energy Physics, Protvino}
\affiliation{Institute for Theoretical and Experimental Physics, Moscow}
\affiliation{J. Stefan Institute, Ljubljana}
\affiliation{Kanagawa University, Yokohama}
\affiliation{Institut f\"ur Experimentelle Kernphysik, Karlsruhe Institut f\"ur Technologie, Karlsruhe}
\affiliation{Korea Institute of Science and Technology Information, Daejeon}
\affiliation{Korea University, Seoul}
\affiliation{Kyungpook National University, Taegu}
\affiliation{\'Ecole Polytechnique F\'ed\'erale de Lausanne (EPFL), Lausanne}
\affiliation{Faculty of Mathematics and Physics, University of Ljubljana, Ljubljana}
\affiliation{University of Maribor, Maribor}
\affiliation{Max-Planck-Institut f\"ur Physik, M\"unchen}
\affiliation{University of Melbourne, School of Physics, Victoria 3010}
\affiliation{Nagoya University, Nagoya}
\affiliation{Nara Women's University, Nara}
\affiliation{National Central University, Chung-li}
\affiliation{National United University, Miao Li}
\affiliation{Department of Physics, National Taiwan University, Taipei}
\affiliation{H. Niewodniczanski Institute of Nuclear Physics, Krakow}
\affiliation{Nippon Dental University, Niigata}
\affiliation{Niigata University, Niigata}
\affiliation{University of Nova Gorica, Nova Gorica}
\affiliation{Novosibirsk State University, Novosibirsk}
\affiliation{Panjab University, Chandigarh}
\affiliation{Peking University, Beijing}
\affiliation{Saga University, Saga}
\affiliation{University of Science and Technology of China, Hefei}
\affiliation{Seoul National University, Seoul}
\affiliation{Sungkyunkwan University, Suwon}
\affiliation{School of Physics, University of Sydney, NSW 2006}
\affiliation{Tata Institute of Fundamental Research, Mumbai}
\affiliation{Excellence Cluster Universe, Technische Universit\"at M\"unchen, Garching}
\affiliation{Toho University, Funabashi}
\affiliation{Tohoku Gakuin University, Tagajo}
\affiliation{Tohoku University, Sendai}
\affiliation{Department of Physics, University of Tokyo, Tokyo}
\affiliation{Tokyo Metropolitan University, Tokyo}
\affiliation{Tokyo University of Agriculture and Technology, Tokyo}
\affiliation{IPNAS, Virginia Polytechnic Institute and State University, Blacksburg, Virginia 24061}
\affiliation{Yonsei University, Seoul}
  \author{A.~Poluektov}\affiliation{Budker Institute of Nuclear Physics, Novosibirsk}\affiliation{Novosibirsk State University, Novosibirsk} 
  \author{A.~Bondar}\affiliation{Budker Institute of Nuclear Physics, Novosibirsk}\affiliation{Novosibirsk State University, Novosibirsk} 
  \author{B.~D.~Yabsley}\affiliation{School of Physics, University of Sydney, NSW 2006} 
  \author{I.~Adachi}\affiliation{High Energy Accelerator Research Organization (KEK), Tsukuba} 
  \author{H.~Aihara}\affiliation{Department of Physics, University of Tokyo, Tokyo} 
  \author{K.~Arinstein}\affiliation{Budker Institute of Nuclear Physics, Novosibirsk}\affiliation{Novosibirsk State University, Novosibirsk} 
  \author{V.~Aulchenko}\affiliation{Budker Institute of Nuclear Physics, Novosibirsk}\affiliation{Novosibirsk State University, Novosibirsk} 
  \author{T.~Aushev}\affiliation{\'Ecole Polytechnique F\'ed\'erale de Lausanne (EPFL), Lausanne}\affiliation{Institute for Theoretical and Experimental Physics, Moscow} 
  \author{A.~M.~Bakich}\affiliation{School of Physics, University of Sydney, NSW 2006} 
  \author{V.~Balagura}\affiliation{Institute for Theoretical and Experimental Physics, Moscow} 
% ...
\collaboration{The Belle Collaboration}

\begin{abstract} 

We present a new measurement of the unitarity triangle angle $\phi_3$ 
using a Dalitz plot analysis of the $K^0_S\pi^+\pi^-$ decay of the neutral $D$ meson 
produced in $B^{\pm}\to D^{(*)}K^{\pm}$ decays. The method exploits the interference between 
$D^0$ and $\overline{D}{}^0$ to extract the angle $\phi_3$, strong phase 
$\delta$ and the ratio $r$ of suppressed and allowed amplitudes. We apply 
this method to a 605 fb$^{-1}$ data sample collected by the Belle experiment. 
The analysis uses three decays: $B^{\pm}\to DK^{\pm}$, and $B^{\pm}\to D^{*}K^{\pm}$ with 
$D^{*}\to D\pi^0$ and $D^{*}\to D\gamma$, as well as the corresponding charge-conjugate modes. 
From a combined maximum likelihood fit to the three modes, we obtain 
$\phi_3=78.4^{\circ}{}^{+10.8^{\circ}}_{-11.6^{\circ}}\pm 3.6^{\circ}(\mbox{syst})
\pm 8.9^{\circ}(\mbox{model})$. 
$CP$ conservation in this process is ruled out at the 
confidence level $(1-CL)=5\times 10^{-4}$, or 3.5 standard deviations.
\end{abstract}
\maketitle

\tighten

\section{Introduction}

Determinations of parameters of the Cabibbo-Kobayashi-Maskawa
(CKM) matrix \cite{ckm} are important as checks on the consistency of the
Standard Model, and as ways to search for new physics. Among the
angles of the CKM unitarity triangle, $\phi_3$ (also widely known as $\gamma$) 
is the least-well
constrained by direct measurements, so new results are of
particular interest. The principal experimental resource is $CP$
violation in the family of decays $B \to D K$: various methods for
extracting a $\phi_3$ measurement have been proposed
\cite{glw,dunietz,eilam,ads}, following the original discussion of
direct $CP$ violation measurement by Bigi, Carter, and Sanda~\cite{bigi}. 
The most sensitive technique relies
on three-body final states~\cite{giri,binp_dalitz} such as \kspp.

In the Wolfenstein parameterization of the CKM matrix elements~\cite{wolfenstein}, 
the weak parts of the amplitudes that contribute to the decay \bdkp\ 
are given by $V_{cb}^*V_{us\vphantom{b}}^{\vphantom{*}}\sim A\lambda^3$ 
(for the $\db K^+$ final state) and
$V_{ub}^*V_{cs\vphantom{b}}^{\vphantom{*}}\sim A\lambda^3(\rho+i\eta)$ 
(for $\dn K^+$); the two amplitudes interfere as the \dn and \db 
mesons decay into the same final state \kspp. 
Assuming no $CP$ asymmetry in neutral $D$ decays, 
the amplitude 
for the process $B^{\pm}\to (K^0_S\pi^+\pi^-)_{D}K^{\pm}$
as a function of the Dalitz plot variables $m^2_+=m^2_{K^0_S\pi^+}$ and 
$m^2_-=m^2_{K^0_S\pi^-}$ is 
\begin{equation}
  M_{\pm}=f(m^2_{\pm}, m^2_{\mp})+re^{\pm i\phi_3+i\delta}f(m^2_{\mp}, m^2_{\pm}), 
\end{equation}
where $f(m^2_+, m^2_-)$ is the amplitude of the \dkpp decay,
$r$ is the ratio of the magnitudes of the two interfering amplitudes, 
and $\delta$ is the strong phase difference between them. 
The \dkpp decay amplitude $f$ can be determined
from a large sample of flavor-tagged \dkpp decays 
produced in continuum $e^+e^-$ annihilation. Once $f$ is known, 
a simultaneous fit to $B^+$ and $B^-$ data allows the 
contributions of $r$, $\phi_3$ and $\delta$ to be separated. 
The method has a two-fold ambiguity: 
$(\phi_3,\delta)$ and $(\phi_3+180\deg, \delta+180\deg)$
solutions cannot be separated. We always choose the solution 
with $0<\phi_3<180\deg$. We neglect the effects of charm mixing in 
this formalism. 
Given the current precision of $\phi_3$ and the constraints 
on the $D^0$ mixing parameters
($x_D, y_D\sim 0.01$ \cite{mixing_hfag}),
these effects can be safely neglected \cite{mixing_phi3}, although
it is possible to take them into account if they appear
to be significant for future precision measurements.
References \cite{giri} and \cite{belle_phi3_3} give  
a more detailed description of the technique. 

This method can be applied to other decay modes: in addition 
to \bdkp,\footnote{Charge-conjugate modes are implied 
throughout the paper unless noted otherwise.} excited states of neutral $D$ and $K$
can also be used, although 
the values of $\delta$ and $r$ can differ for these decays. 
Both the BaBar and Belle collaborations have successfully applied this 
technique to \bddsks modes with \dn decaying to 
\kspp~\cite{babar_phi3_2, babar_phi3_3, belle_phi3_1,
belle_phi3_2, belle_phi3_3}. 
In addition, the BaBar collaboration reported a measurement of 
$\phi_3$ using the \bdk mode with the $D$ decaying to the 
$K^0_S K^+K^-$~\cite{babar_phi3_3} and 
$\pi^0\pi^+\pi^-$~\cite{babar_3pi} final states. 

Here we present a measurement of $\phi_3$ using 
the modes \bdkp and \bdskp with \dtkpp, based on a 605 fb$^{-1}$ data 
sample ($657\times 10^6$ $B\overline{B}$ pairs) collected by the Belle 
detector at the KEKB asymmetric $e^+e^-$ collider. 
The Belle detector is described in detail elsewhere \cite{belle,svd2}. 
It is a large-solid-angle magnetic spectrometer consisting of a
silicon vertex detector (SVD), a 50-layer central drift chamber (CDC) for
charged particle tracking and specific ionization measurement ($dE/dx$), 
an array of aerogel threshold Cherenkov counters (ACC), time-of-flight
scintillation counters (TOF), and an array of CsI(Tl) crystals for 
electromagnetic calorimetry (ECL) located inside a superconducting solenoid coil
that provides a 1.5 T magnetic field. An iron flux return located outside 
the coil is instrumented to detect $K_L$ mesons and identify muons (KLM).

The results presented in this paper supersede our previous measurement 
based on a sample of 
$386\times 10^6$ $B\overline{B}$ pairs~\cite{belle_phi3_3}.
In addition to \bdk and the \bdsk mode with \dsndpi, this analysis 
exploits \bdsk with \dsndg. 
The \dsndg mode has nearly the same parameters as 
\bdsk with \dsndpi, the only difference being that due to the opposite 
$C$ parities of the $\gamma$ and $\pi^0$, the strong phases 
for these modes differ by $180^{\circ}$~\cite{bondar_gershon}. 
This provides an additional cross-check for the analysis and 
allows systematic uncertainties in the combined measurement to be reduced. 
The analysis procedure is also improved. 
It uses additional variables in the maximum likelihood fit
for the separation of signal from background; this allows
one to relax some selection requirements, thus increasing
the sample size.

\section{Event selection}

\label{event_selection}

The decay chains \bdkp and \bdskp with \dsndpi
and \dsndg are selected for the analysis. The neutral $D$ meson 
is reconstructed in the \kspp final state in all cases. 
We also select decays \dsdpim produced via the 
$e^+e^-\to c\bar{c}$ continuum process as a high-statistics 
sample to determine the \dkpp decay amplitude. 

Charged tracks are required to satisfy criteria based on the 
quality of the track fit and the distance from the interaction point. 
We require each track to have a transverse momentum greater than 
100 MeV/$c$. (The reference axis is given by the direction of the 
$e^+$ beam.) Separation of kaons and pions is accomplished by combining 
the responses of 
the ACC and the TOF with the $dE/dx$ measurement from the CDC. 
Photon candidates are required to have ECL energy greater than 30~MeV. 
Neutral pion candidates are formed from pairs of photons with invariant 
masses in the range 120~MeV/$c^2$ to 150~MeV/$c^2$. 
Neutral kaons are reconstructed from pairs of oppositely charged tracks
with an invariant mass $M_{\pi\pi}$ within $7$~MeV/$c^2$ of the nominal 
$K^0_S$ mass, and forming a vertex more than 1~mm from the interaction
point in the transverse plane.

To determine the \dkpp\ decay amplitude we use $D^{*\pm}$ mesons
produced via the $e^+ e^-\to c\bar{c}$ continuum process. 
The flavor of the neutral $D$ meson is tagged by the charge of the slow pion 
(which we denote as $\pi_s$) in the decay \dsdpims.
The slow pion track is fitted to the $\overline{D}{}^0$ 
production vertex to 
improve the momentum and angular resolution of the $\pi_s$. 
To select neutral $D$ candidates we require the invariant mass of the 
$K^0_S\pi^+\pi^-$ system to be within 11 MeV/$c^2$ of the $D^0$ mass.
To select events originating from a $D^{*-}$ decay 
we impose a requirement on the difference 
$\Delta M$ of the invariant 
masses of the $D^{*-}$ and the neutral $D$ candidates: 
$144.9\mbox{ MeV}/c^2<\Delta M<145.9\mbox{ MeV}/c^2$.
Suppression of the combinatorial background from $B\overline{B}$ events
is achieved by requiring the $D^{*-}$ momentum 
in the center-of-mass (CM) frame to be greater than 3.0~GeV/$c$.
The number of events in the signal region is $290.9\times 10^3$; 
the background fraction is 1.0\%.

The selection of $B$ candidates is based on the CM energy difference
$\de = \sum E_i - E_{\rm beam}$ and the beam-constrained $B$ meson mass
$\mbc = \sqrt{E_{\rm beam}^2 - (\sum \vec{p}_i)^2}$, where $E_{\rm beam}$ 
is the CM beam 
energy, and $E_i$ and $\vec{p}_i$ are the CM energies and momenta of the
$B$ candidate decay products. We select events with $\mbc>5.2$ GeV/$c^2$
and $|\de|<0.15$~GeV for further analysis. 
We also impose a requirement on the invariant mass of the neutral $D$ 
candidate as above: 
$|M_{\kspp}-M_{D^0}|<11$~MeV/$c^2$. To obtain the Dalitz plot variables 
$m^2_+$ and $m^2_-$, a kinematical fit is employed with the constraint
that the $\kspp$ invariant mass be equal to $M_{D^0}$.

We consider two major background sources in our data: the continuum 
process $e^+e^-\to q\bar{q}$, where the light component with $q=u, d, s$ 
and the charmed component are treated separately; and $B\overline{B}$
decays, where events with real $D^0$ (due to \bdpi etc.) are treated 
separately. To suppress background from continuum events, we calculate two 
variables that characterize the 
event shape. One is the cosine of the thrust angle \thr, 
where $\theta_{\rm thr}$ is the angle between the thrust axis of 
the $B$ candidate daughters and that of the rest of the event, 
calculated in the CM frame. 
The other is a Fisher discriminant \fish composed of 11 parameters \cite{fisher}: 
the production angle of the $B$ candidate, the angle of the $B$ thrust 
axis relative to the beam axis, and nine parameters representing 
the momentum flow in the event relative to the $B$ thrust axis in the CM frame.
In the first stage of the analysis, the $(\mbc, \de)$
distribution is fitted in order to obtain the fractions of the 
background components, and we require $|\thr|<0.8$ and $\fish>-0.7$. 
In the Dalitz plot fit, we do not reject events based on these 
variables (as in the previous analysis~\cite{belle_phi3_3}), 
but rather use them in the likelihood function to 
better separate signal and background events. 
This leads to a 7--8\% improvement in the expected statistical error. 

\begin{figure}

  \epsfig{figure=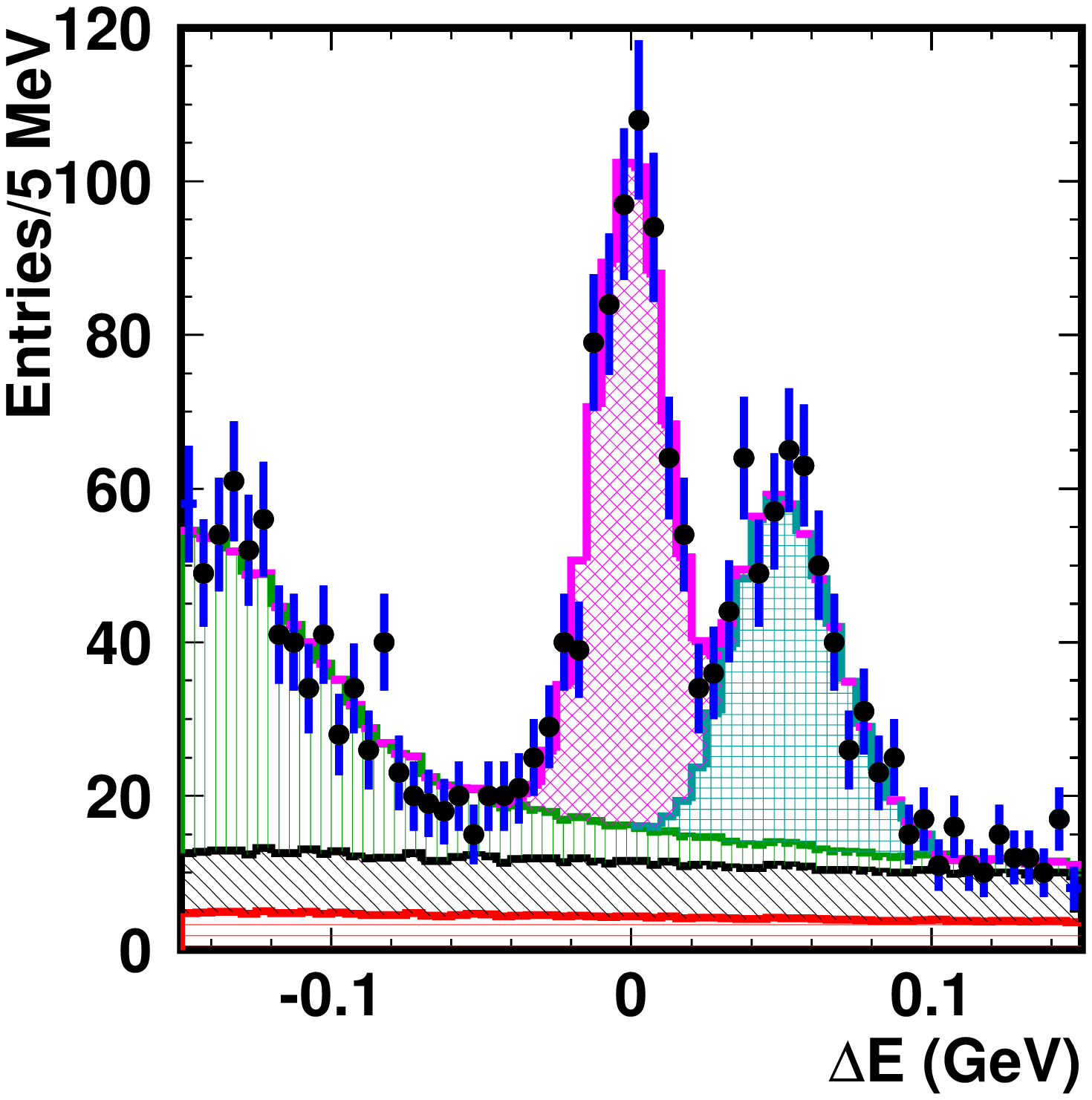,width=0.23\textwidth}
  \put(-25,95){\mbox{(a)}}
  \epsfig{figure=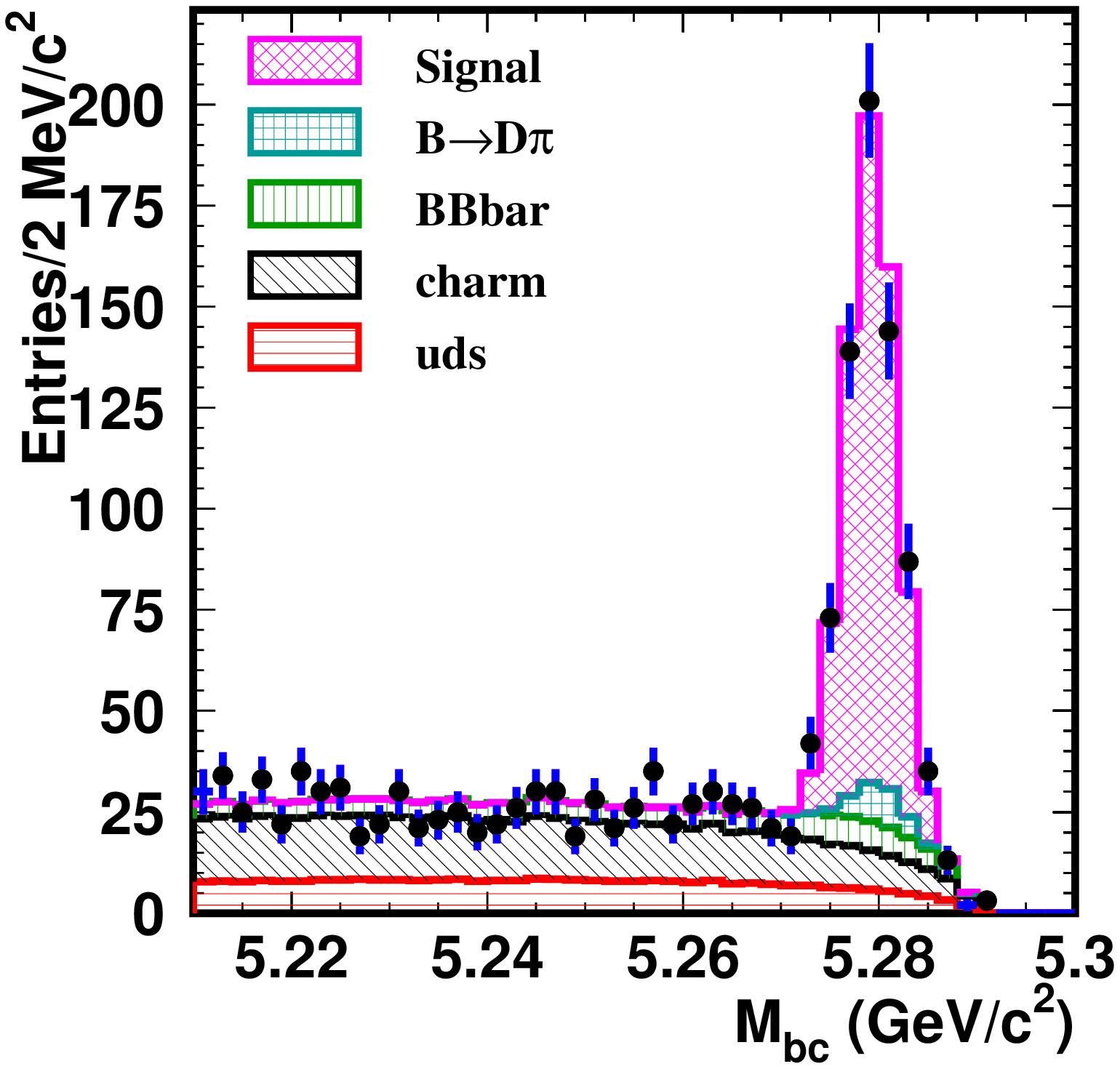,width=0.23\textwidth}
  \put(-25,95){\mbox{(b)}}

  \epsfig{figure=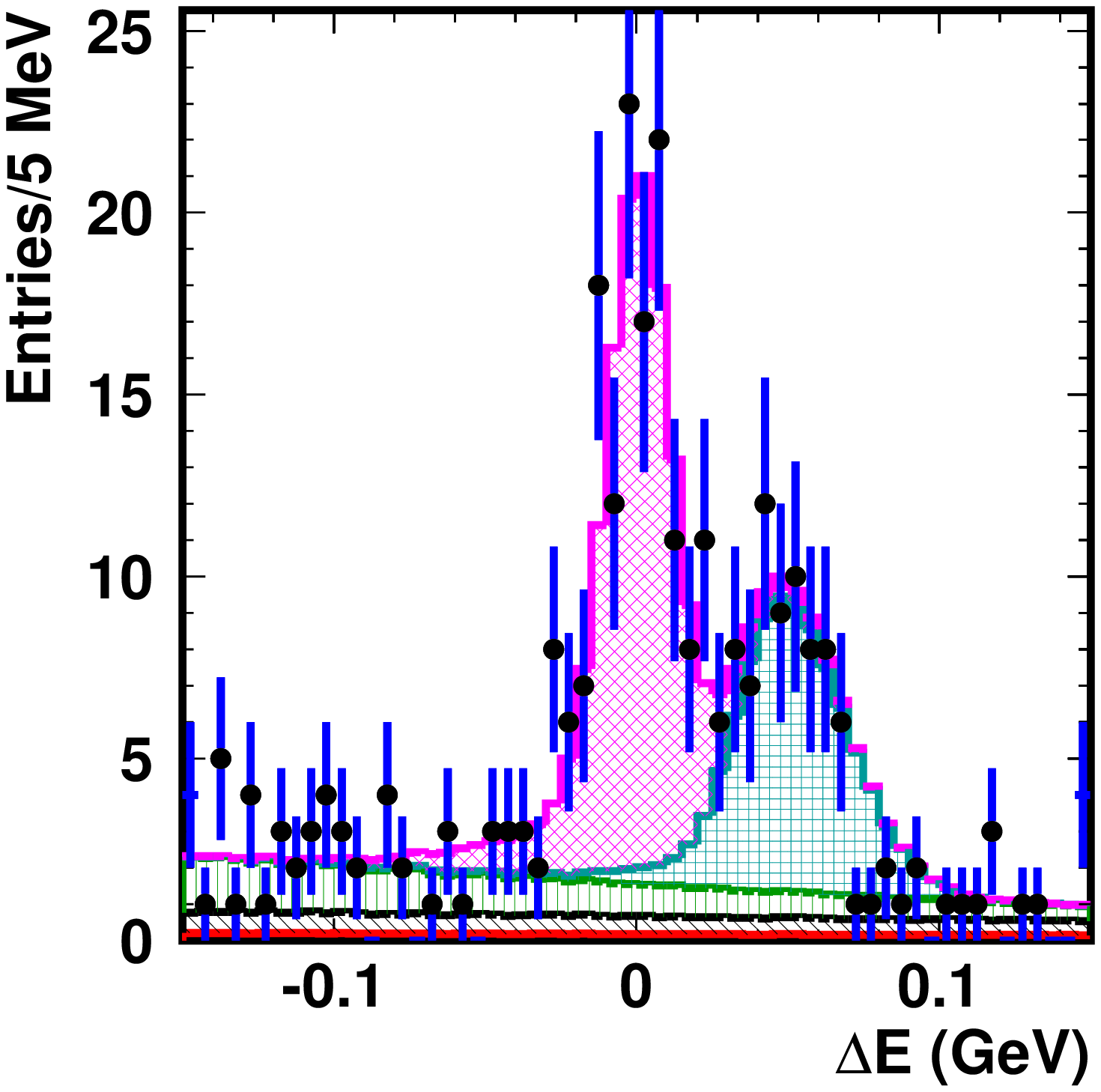,width=0.23\textwidth}
  \put(-25,95){\mbox{(c)}}
  \epsfig{figure=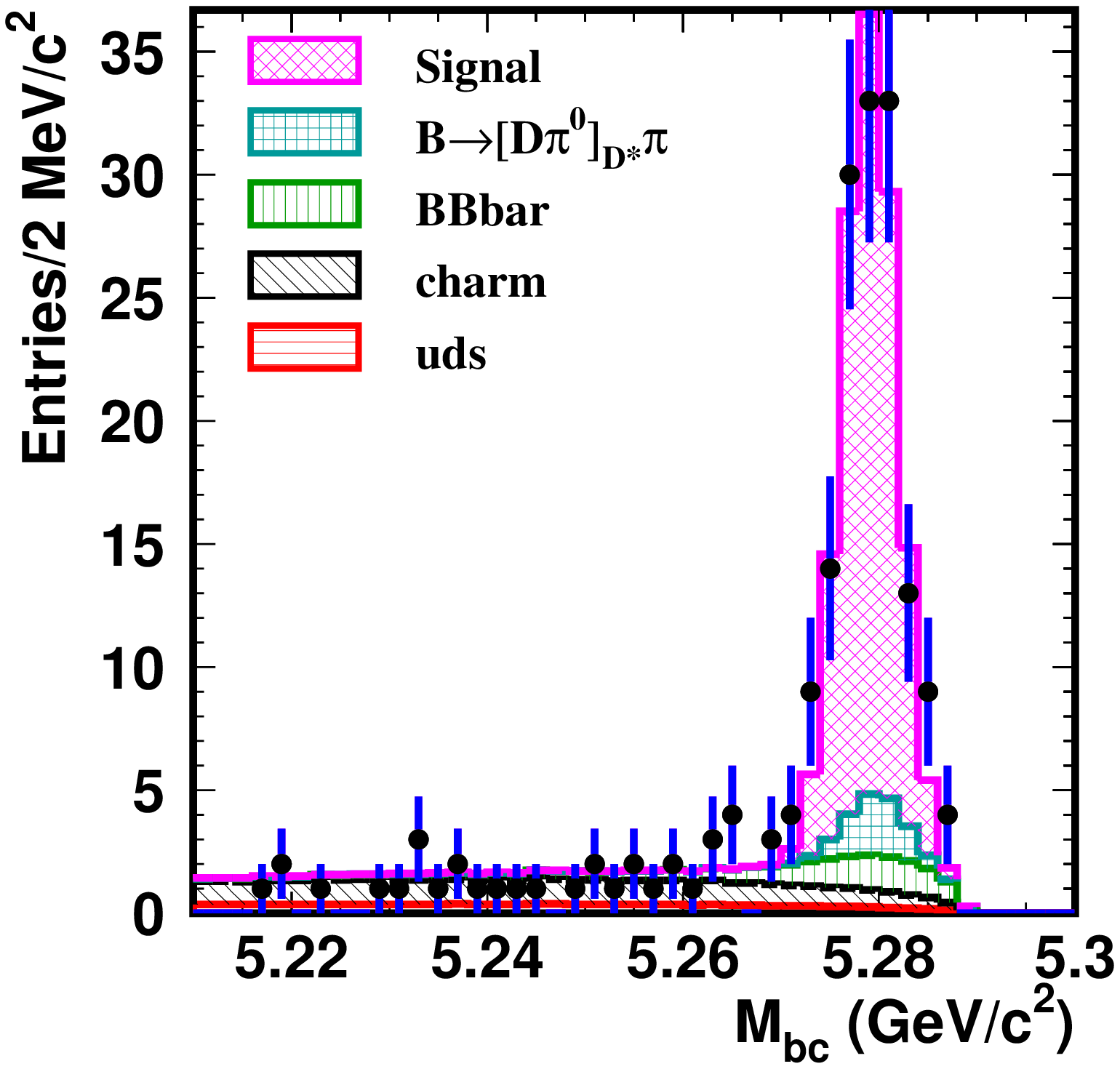,width=0.23\textwidth}
  \put(-25,95){\mbox{(d)}}

  \epsfig{figure=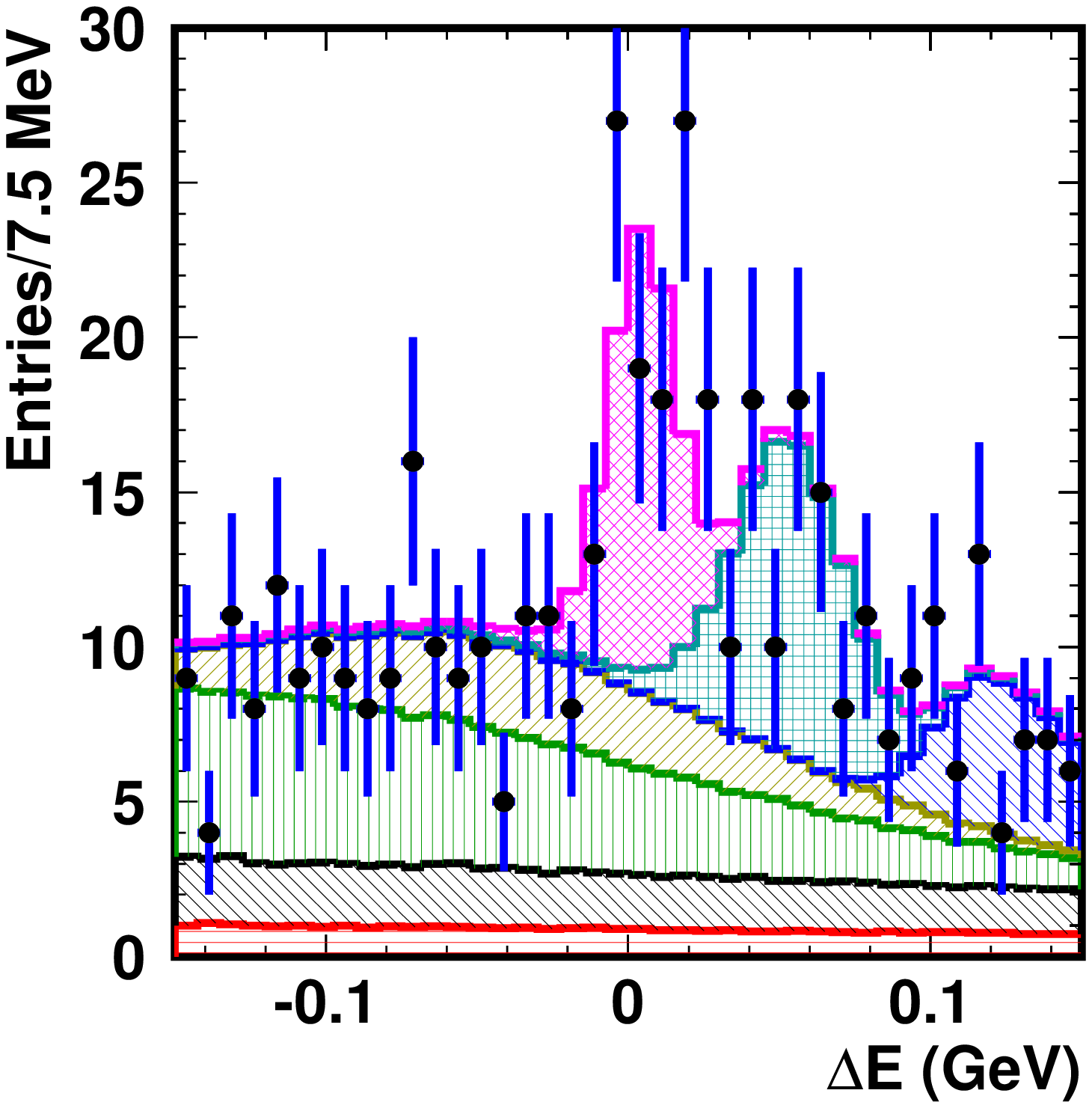,width=0.23\textwidth}
  \put(-25,95){\mbox{(e)}}
  \epsfig{figure=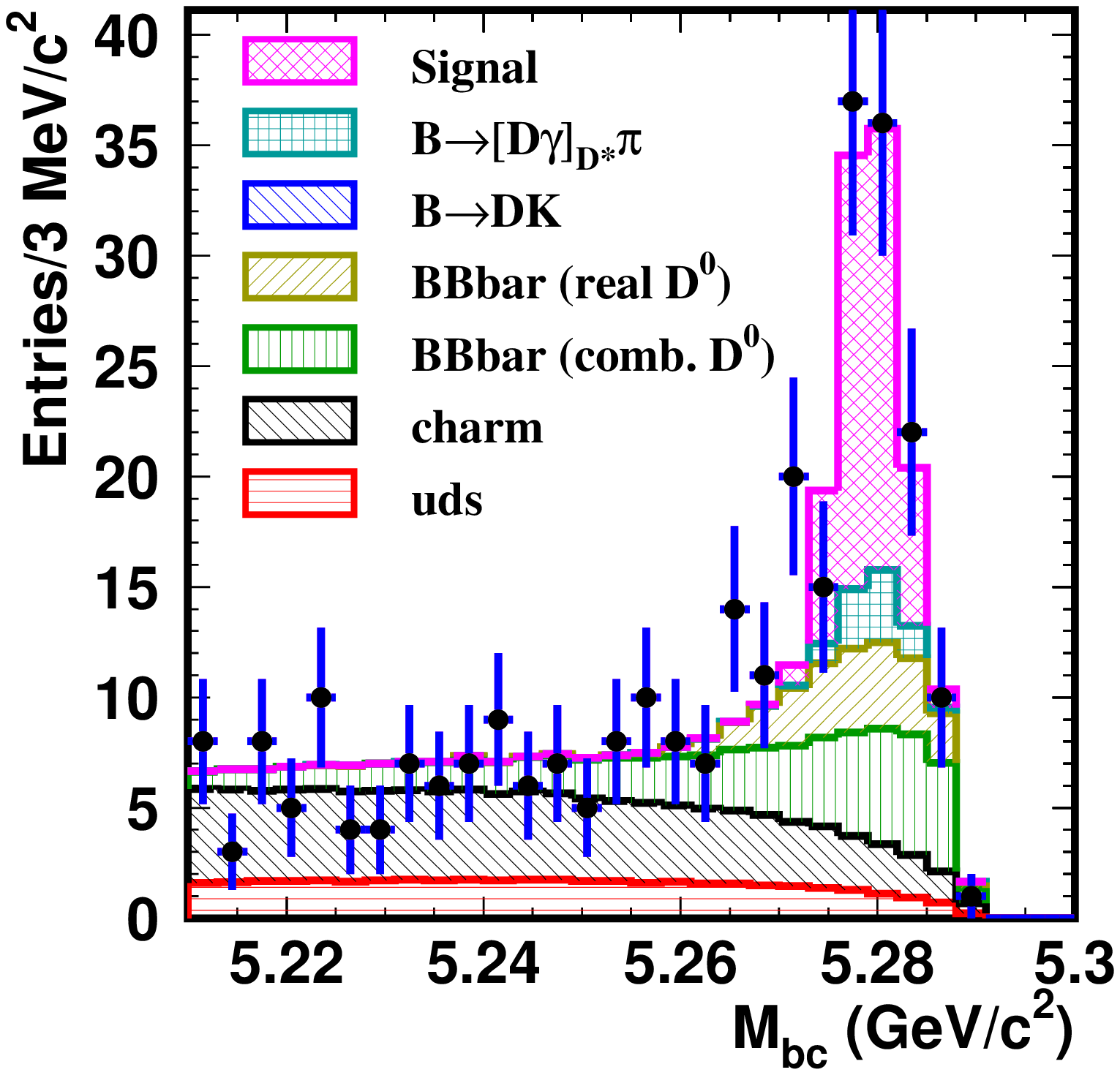,width=0.23\textwidth}
  \put(-25,95){\mbox{(f)}}
 
  \caption{$\Delta E$ and $M_{\rm bc}$ distributions for the 
           \bdkp (a,b), \bdskp with \dsndpi (c,d), and 
           \bdskp with \dsndg (e,f) event samples. 
           Points with error bars are the data, and
           the histograms are fitted contributions due to signal, 
           misidentified \bddspi events, and $B\overline{B}$, 
           charm, and $(u,d,s)$ backgrounds; in
           (e), a \bdk contribution with random photon is also
           included.
           $\Delta E$ distributions are plotted with a
           $M_{\rm bc}>5.27$ MeV/$c^2$ requirement;  
           $M_{\rm bc}$ distributions use 
           a $|\Delta E|<30$ MeV requirement. $|\thr|<0.8$
           and $\fish>-0.7$ requirements are used in all 
           the plots. 
           }
  \label{signal_mbcde}
\end{figure}

\begin{figure}

  \epsfig{figure=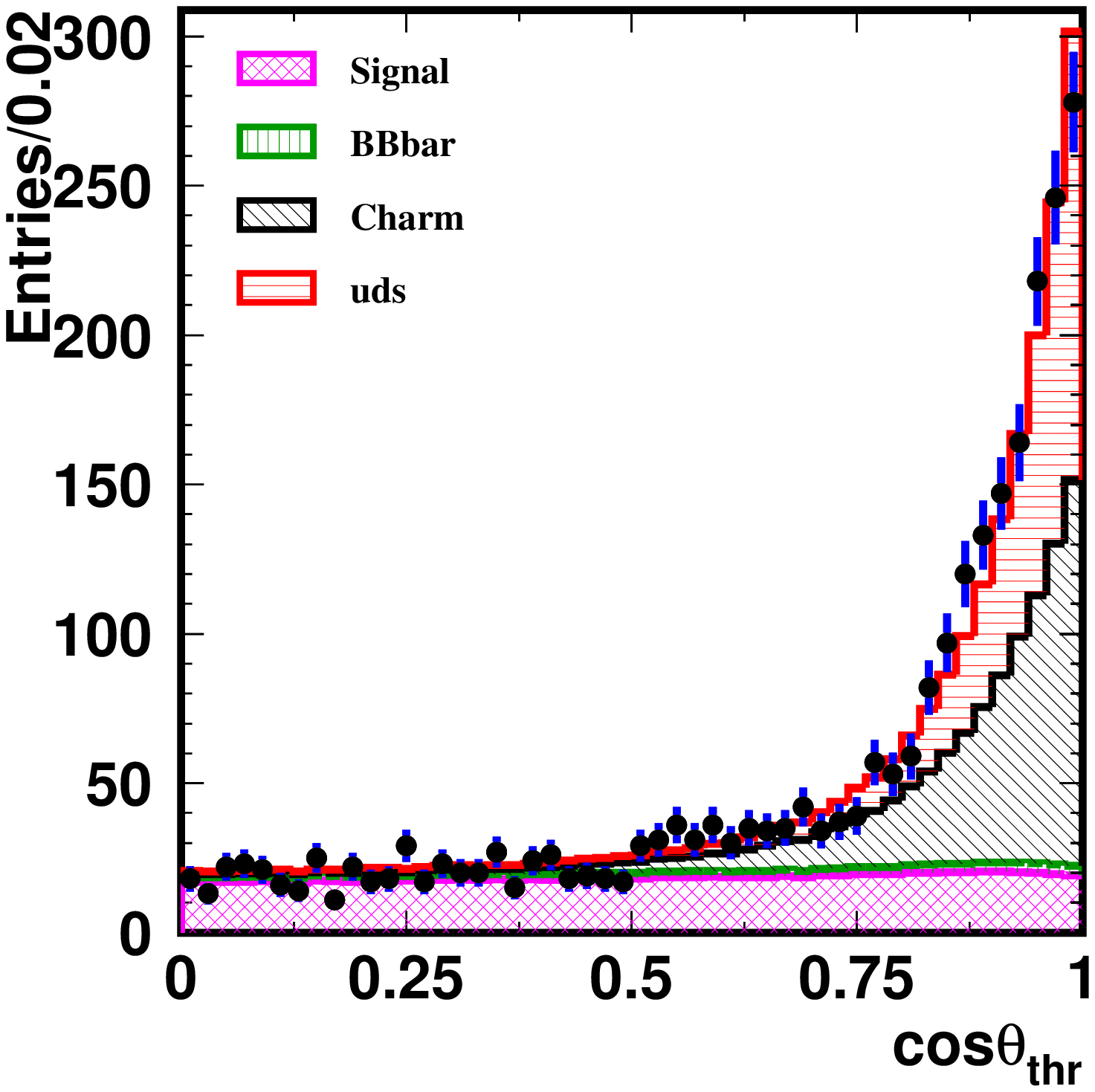,width=0.23\textwidth}
  \put(-25,95){\mbox{(a)}}
  \epsfig{figure=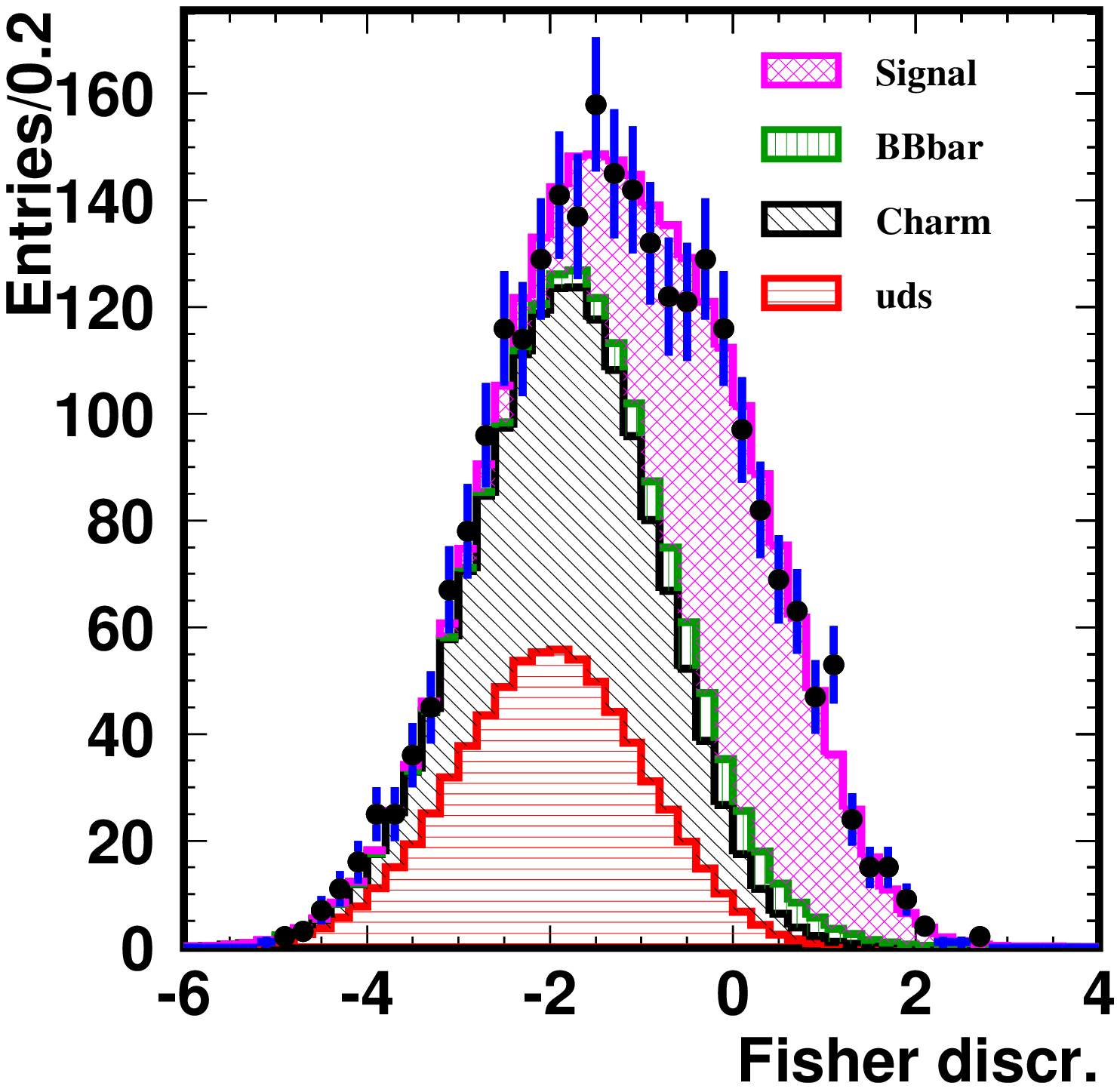,width=0.23\textwidth}
  \put(-25,75){\mbox{(b)}}

  \caption{$\thr$ and $\fish$ distributions for the 
           \bdkp event sample. 
           Points with error bars show the data with 
           $M_{\rm bc}>5.27$ MeV/$c^2$ and 
           $|\Delta E|<30$ MeV requirements, and
           the histograms are fitted contributions due to signal, 
           $B\overline{B}$, charm, and $(u,d,s)$ backgrounds. 
           }
  \label{signal_thrfish}
\end{figure}

The \de and \mbc distributions for the \bdkp
mode are shown in Fig.~\ref{signal_mbcde}~(a), (b). 
For the selected events a two-dimensional unbinned maximum 
likelihood fit in the variables \mbc and \de 
is performed, with the fractions of continuum, $B\bar{B}$ and \bddspi backgrounds
as free parameters, and their distributions fixed from generic MC 
simulation. (The continuum component is also split into 
$(u,d,s)$ and charm components in the figure, based on fractions in the MC.)
The resulting signal and background fractions
are used in the Dalitz plot fit to obtain the event-by-event signal 
to background ratio. A more detailed description of the two-stage 
procedure is given in Section~\ref{dalitz_analysis}.
The number of events in the signal box 
($\mbc>5.27$ GeV/$c^2$, $|\de|<30$ MeV, $|\thr|<0.8$, $\fish>-0.7$)
is 756, with a signal purity of $(70.5\pm 1.2)$\%. 
The $(\mbc,\de)$ fit yields a continuum background fraction of 
$(17.9\pm 0.7)$\%, a $B\overline{B}$ background fraction 
of $(7.3\pm 0.5)$\%, 
and a \bdpi background fraction of $(4.3\pm 0.3)$\% in the signal box. 
Figure~\ref{signal_thrfish} shows the distributions of 
$\thr$ and $\fish$ variables in the $\mbc$, $\de$ signal region for 
the \bdkp mode. The distributions for the other modes are similar.

To select \bdskp events with \dsndpi, in addition to the 
requirements described above, we require that the mass difference 
$\Delta M$ of neutral $D^{*}$ and $D$ candidates satisfies 
$140\mbox{ MeV}/c^2<\Delta M<144\mbox{ MeV}/c^2$. 
The \de and \mbc distributions for this mode are shown in 
Fig.~\ref{signal_mbcde} (c), (d). 
The background fractions are obtained in the same way as for \bdk mode.
The number of events in the signal box is 149, with $(79.7\pm 2.5)$\%
signal purity. 
The continuum background fraction is $(5.7\pm 0.7)$\%, the 
$B\overline{B}$ background fraction is $(7.6\pm 1.9)$\%, 
and the \bdspi background fraction is $(7.0\pm 1.3)$\%. 

Selection of the \bdskp mode with \dsndg is performed in a similar way. The 
photon candidate is required to have an energy greater than 100 MeV, 
and the mass difference requirement is $\Delta M<152$ MeV$/c^2$. 
Due to the larger number of background sources for this mode, the 
treatment of background differs. The $B\overline{B}$ background 
is subdivided into events with combinatorial $D$, 
studied using a generic MC sample; and those with real neutral
$D$ mesons, for which a dedicated simulation of each component 
is performed.
The fractions of background components are obtained from an unbinned 4D fit 
of the distribution of variables \mbc, \de, \thr, and \fish. 
The relative fractions of $B\overline{B}$ backgrounds 
with a real $D^0$ (except for \bdspi and \bdk) are fixed according to  
their PDG branching ratios~\cite{pdg} and MC efficiencies.
The \de and \mbc distributions for this mode are shown in 
Fig.~\ref{signal_mbcde} (e), (f). 
The number of events in the signal box is 141, 
and the signal purity is $(41.7\pm 3.6)$\%.
The continuum background fraction is $(15.8\pm 1.3)$\%, 
the fraction of $B\overline{B}$ background with combinatorial 
$D^0$ is $(21.3\pm 3.0)$\%, the contribution of \bdspi, $D^*\to D\gamma$
is $(6.5\pm 1.2)$\%, and the fraction of the rest of $B\overline{B}$ 
events with real $D^0$ is $(14.7\pm 1.1)$\%. 

\begin{figure}

  \epsfig{figure=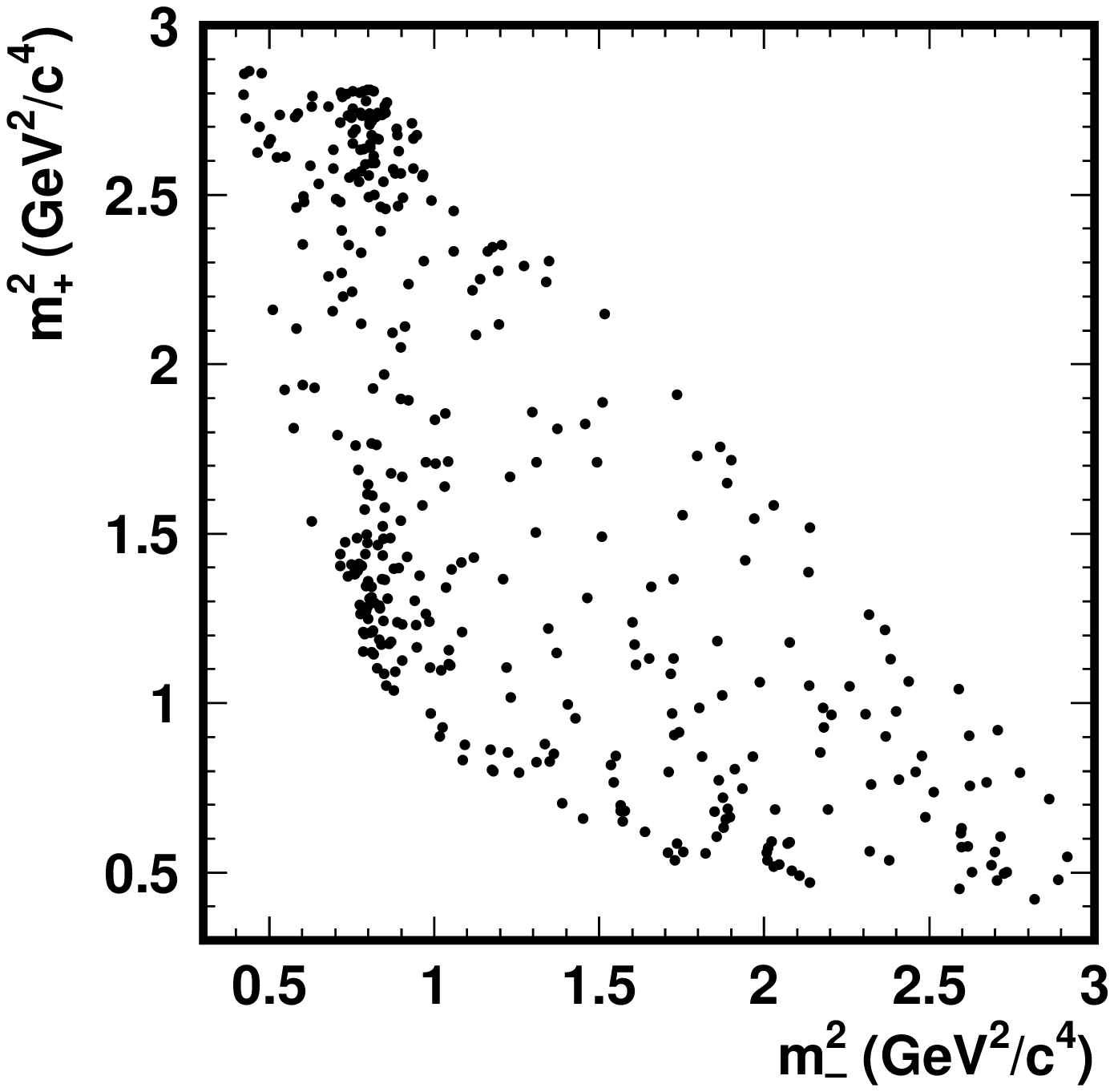,width=0.23\textwidth}
  \put(-25,95){\mbox{(a)}}
  \epsfig{figure=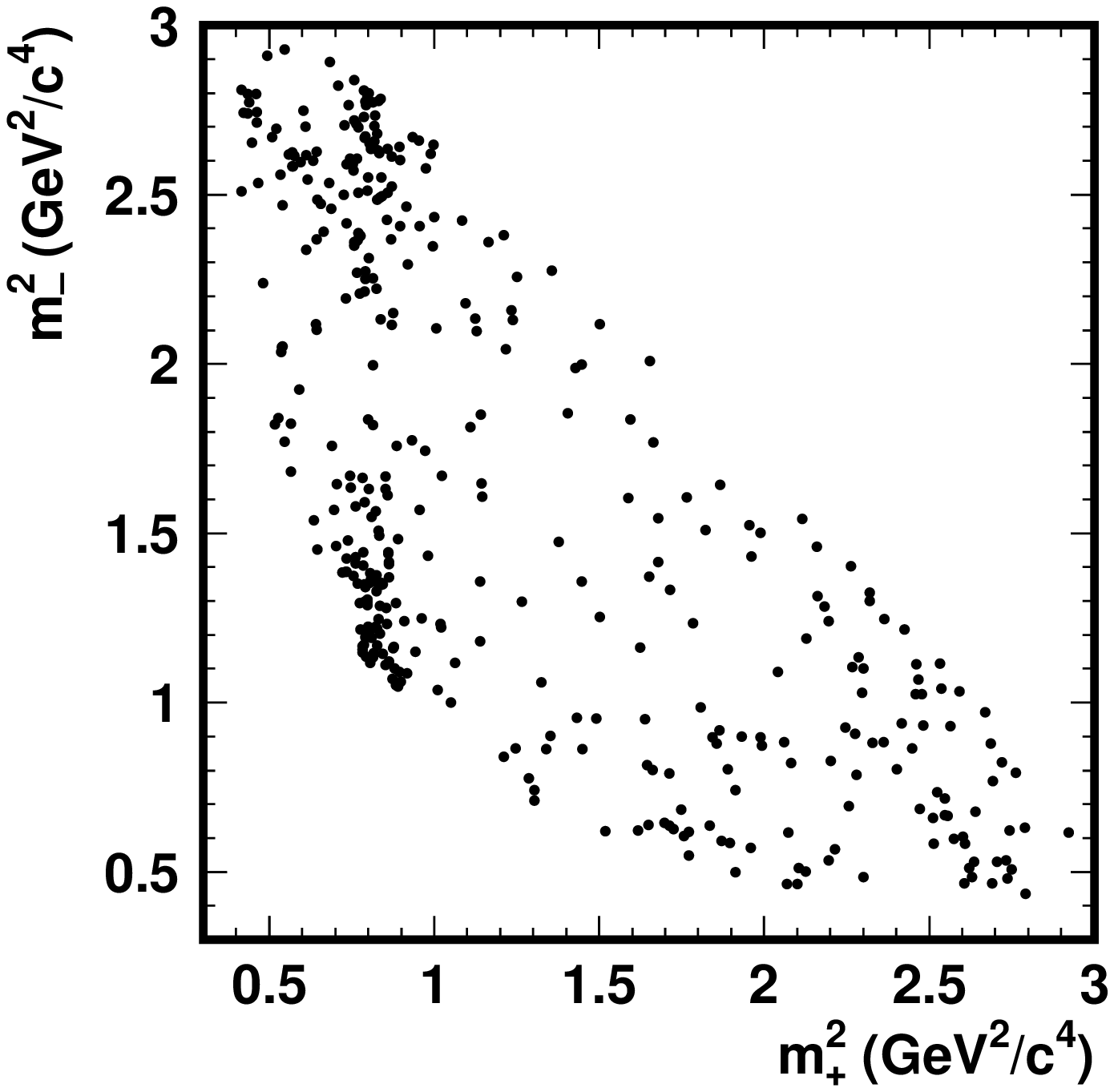,width=0.23\textwidth}
  \put(-25,95){\mbox{(b)}}

  \epsfig{figure=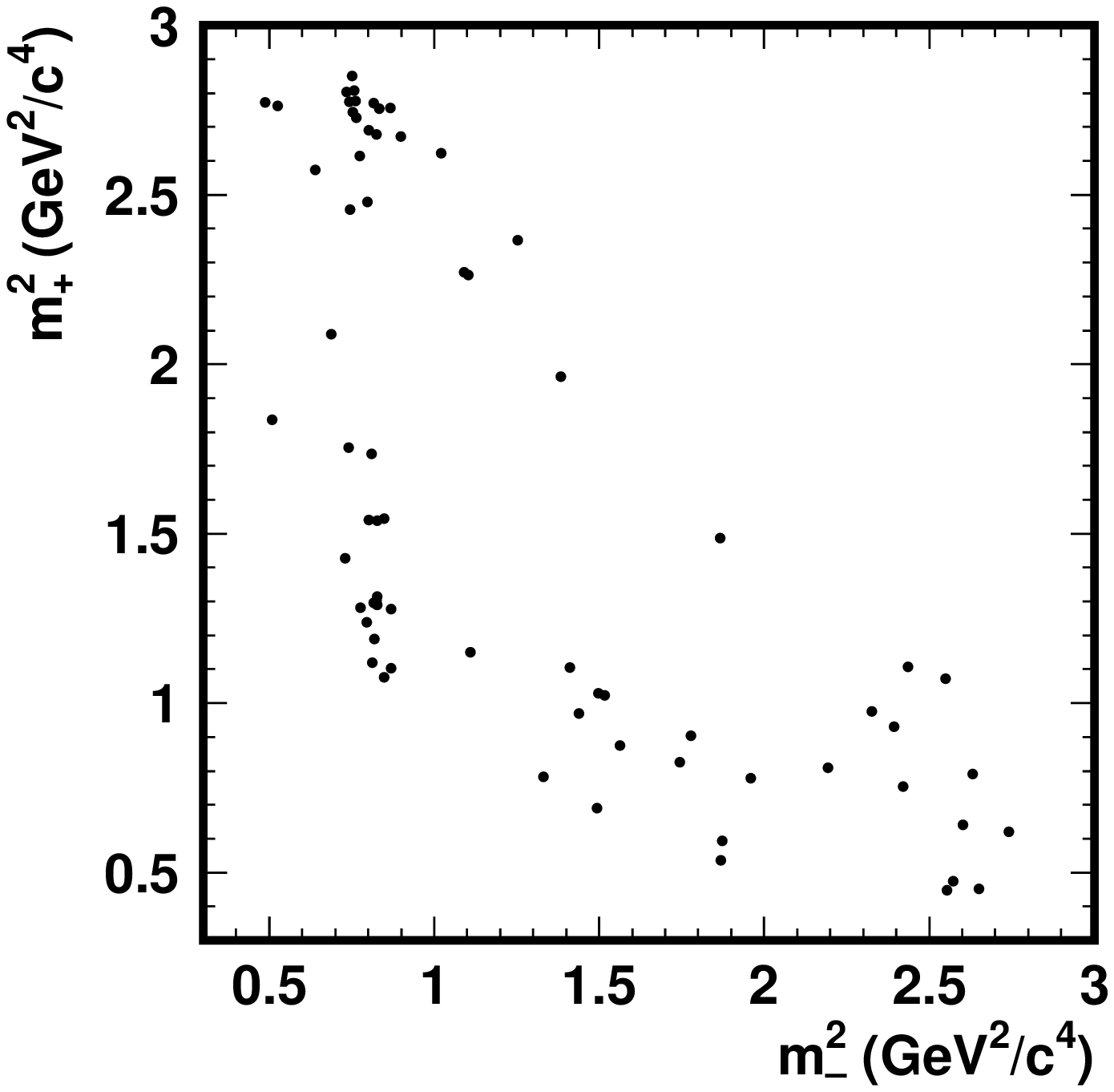,width=0.23\textwidth}
  \put(-25,95){\mbox{(c)}}
  \epsfig{figure=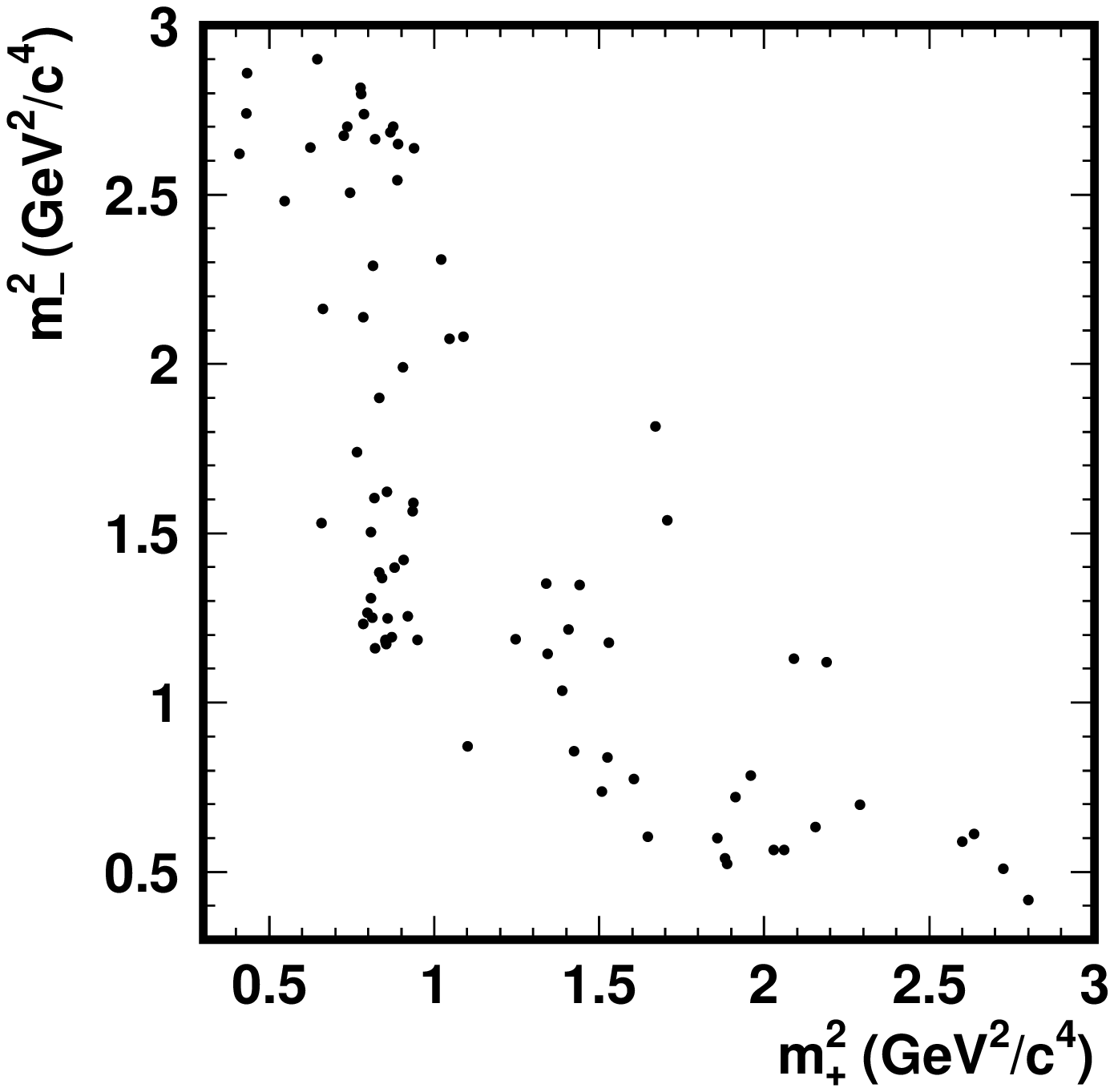,width=0.23\textwidth}
  \put(-25,95){\mbox{(d)}}

  \epsfig{figure=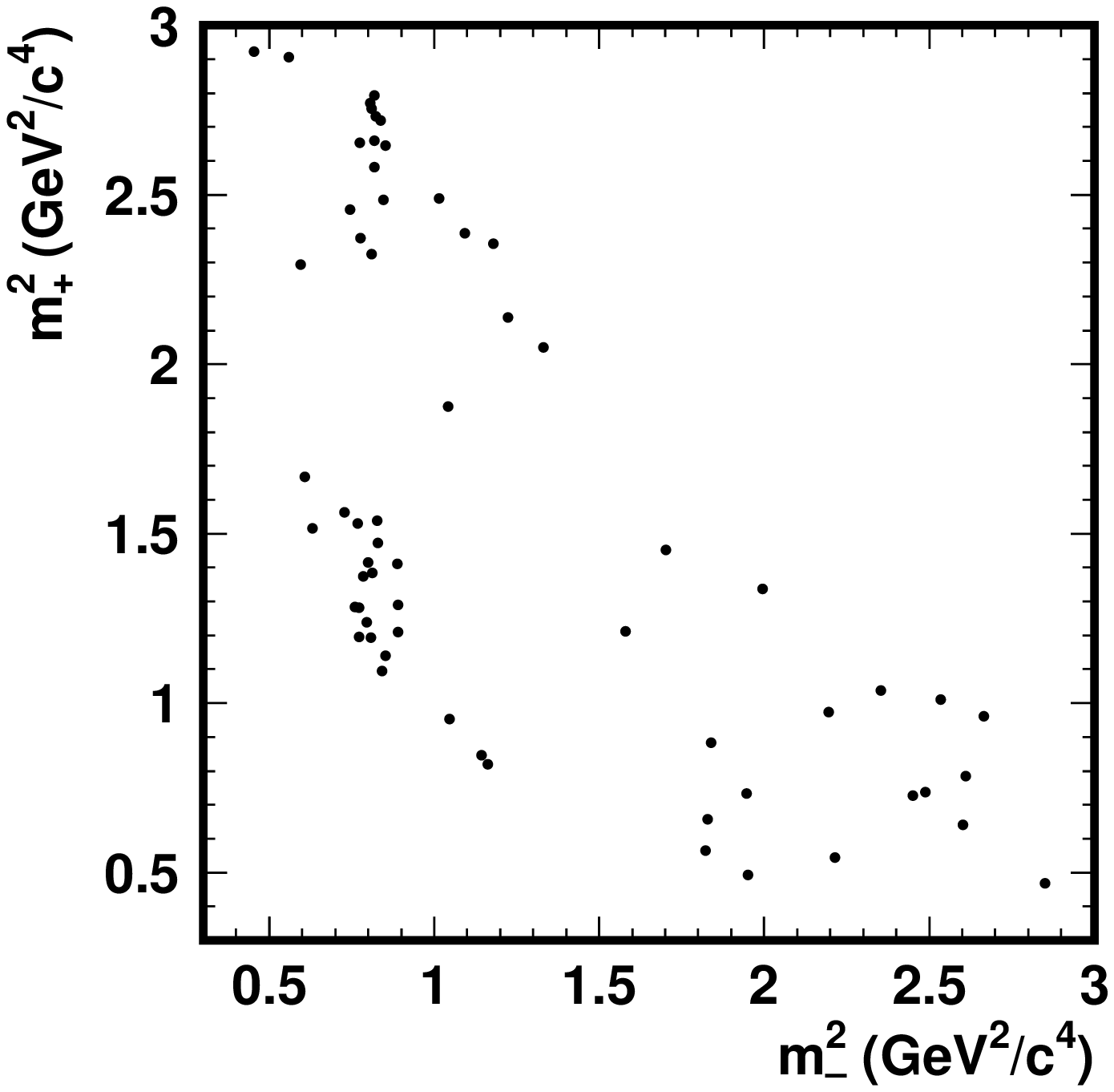,width=0.23\textwidth}
  \put(-25,95){\mbox{(e)}}
  \epsfig{figure=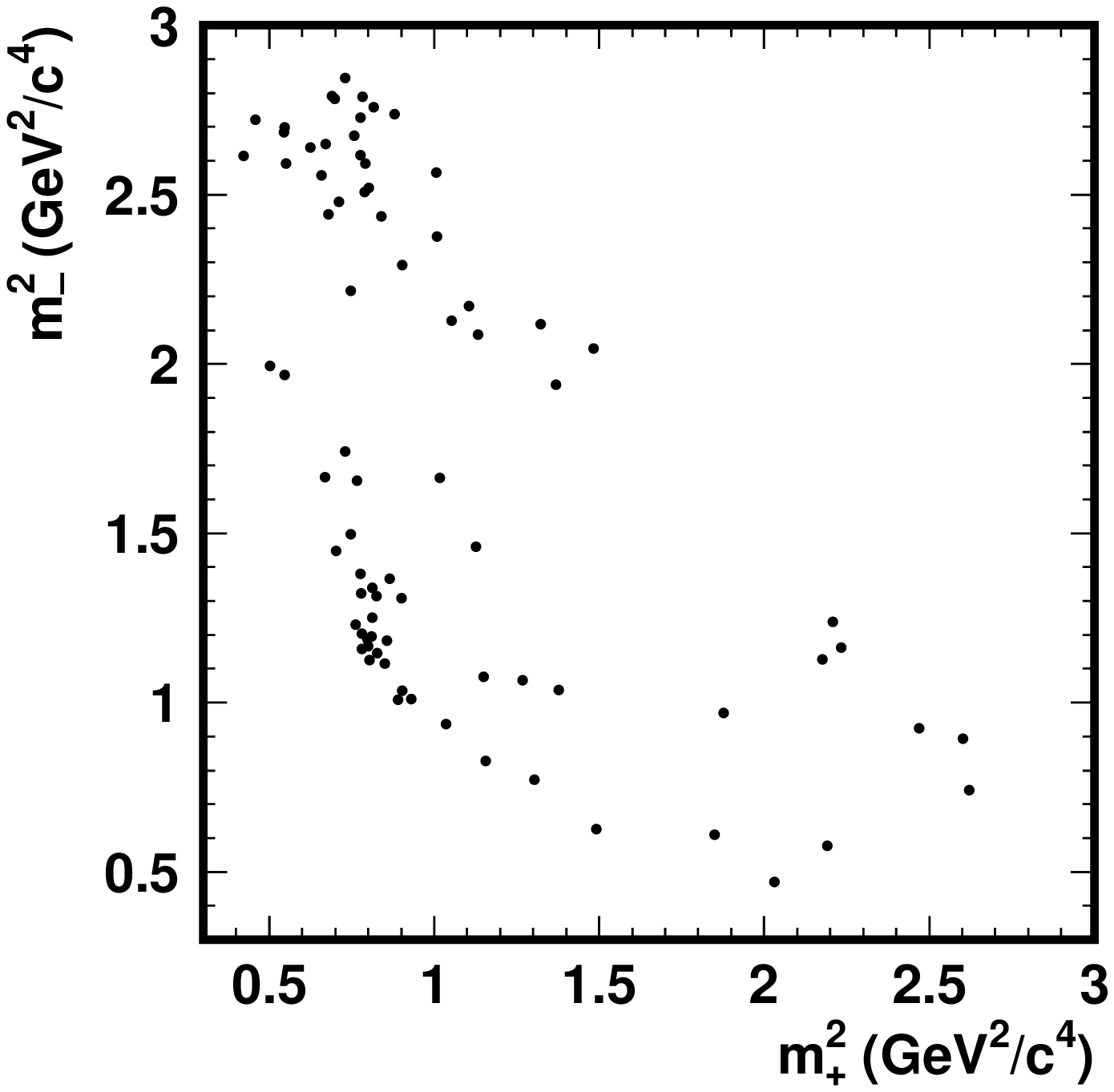,width=0.23\textwidth}
  \put(-25,95){\mbox{(f)}}

  \caption{Dalitz distributions of \dtkpp decays from selected 
            \bdk (a,b), \bdsk with \dsndpi (c,d), and 
            \bdsk with \dsndg (e,f), 
            shown separately for $B^-$ (left) and $B^+$ (right) tags. }
  \label{dalitz_plots}
\end{figure}

The Dalitz distributions of \dtkpp\ decay in the signal box
for each of the \bdk
and \bdsk processes are shown in Fig.~\ref{dalitz_plots}. 

\section{Determination of the \boldmath{\dkpp} decay amplitude}

\label{section_d0_fit}

As in our previous analysis~\cite{belle_phi3_3}, 
the \dkpp decay amplitude is represented using the isobar model. 
The list of resonances 
is also the same, the only difference being the free parameters (mass and 
width) of the $K^*(892)^{\pm}$ and $\rho(770)$ states. 
A modified amplitude, where the scalar $\pi\pi$ component is described 
using the K-matrix approach~\cite{kmatrix}, is used in the estimation of 
the systematic error. 

The amplitude $f$ for the \dkpp decay is described
by a coherent sum of $N$ two-body decay amplitudes and one non-resonant 
decay amplitude,
\begin{equation}
  f(m^2_+, m^2_-) = \sum\limits_{j=1}^{N} a_j e^{i\xi_j}
  \mathcal{A}_j(m^2_+, m^2_-)+
    a_{\text{NR}} e^{i\xi_{\text{NR}}}, 
  \label{d0_model}
\end{equation}
where $\mathcal{A}_j(m^2_+, m^2_-)$ is the matrix element, $a_j$ and 
$\xi_j$ are the amplitude and phase of the matrix element, respectively, 
of the $j$-th resonance, and $a_{\text{NR}}$ and $\xi_{\text{NR}}$
are the amplitude and phase of the non-resonant component. 
The description of the matrix elements follows Ref.~\cite{cleo_model}. 
We use a set of 18 two-body amplitudes. 
These include five Cabibbo-allowed amplitudes: $K^*(892)^+\pi^-$, 
$K^*(1410)^+\pi^-$, $K_0^*(1430)^+\pi^-$, 
$K_2^*(1430)^+\pi^-$ and $K^*(1680)^+\pi^-$;  
their doubly Cabibbo-suppressed partners; and eight amplitudes with
$K^0_S$ and a $\pi\pi$ resonance:
$K^0_S\rho$, $K^0_S\omega$, $K^0_Sf_0(980)$, $K^0_Sf_2(1270)$, 
$K^0_Sf_0(1370)$, $K^0_S\rho(1450)$, $K^0_S\sigma_1$ and $K^0_S\sigma_2$. 

We use an unbinned maximum likelihood technique to fit the Dalitz plot 
distribution to the model described by Eq.~\ref{d0_model} with 
efficiency variation, background contributions and 
finite momentum resolution taken into account. 
The free parameters of the minimization are the amplitudes
$a_j$ and phases $\xi_j$ of the resonances, 
the amplitude $a_{NR}$ and phase $\xi_{NR}$ of the non-resonant component, 
and the masses and widths of the $\sigma_1$ and $\sigma_2$ scalars. 
We also allow the masses and widths of the $K^*(892)^+$ and $\rho(770)$
states to float. 

The procedures for determining the background
density, the efficiency, and the resolution 
are the same as in the previous analyses 
\cite{belle_phi3_2,belle_phi3_3}. 
The background density for \dkpp\ events is extracted from 
$\Delta M$ sidebands. The shape of the efficiency over the Dalitz plot, 
as well as the invariant mass resolution, is extracted from the 
signal Monte-Carlo (MC) simulation. 

The fit results are given in Table~\ref{dkpp_table}. 
The fit fraction for each mode is defined as the ratio 
of the integrals of the squared absolute value of the amplitude for that mode, 
and the squared absolute value of the total amplitude.
The fit fractions do not sum up to unity due to interference effects.
The parameters obtained for the $\sigma_1$ resonance 
($M_{\sigma_1}=(522\pm 6)$~MeV/$c^2$, $\Gamma_{\sigma_1}=(453\pm 10)$~MeV/$c^2$) 
are similar to those found by other experiments~\cite{dkpp_cleo, aitala2}.
The second scalar term $\sigma_2$ is introduced to account for a 
structure observed at $m^2_{\pi\pi} \sim 1.1\,\mathrm{GeV}^2/c^4$:
the fit finds a small but significant contribution with
$M_{\sigma_2}=(1033\pm 7)$~MeV/$c^2$, $\Gamma_{\sigma_2}=(88\pm 7)$~MeV/$c^2$.
Allowing the parameters of the dominant $K^*(892)^+$ and $\rho(770)$
resonances to float results in a significant improvement in the fit quality. 
We obtain $M(K^*(892))=(893.7\pm 0.1)$~MeV/$c^2$, 
$\Gamma(K^*(892))=(48.4\pm 0.2)$~MeV/$c^2$, 
$M(\rho)=(771.7\pm 0.7)$~MeV/$c^2$, and 
$\Gamma(\rho)=(136.0\pm 1.3)$~MeV/$c^2$. 

We perform a $\chi^2$ test using 54$\times$54 bins in the 
region bounded by $m^2_{\pm}=0.3$ GeV$^2/c^4$ and 3.0 GeV$^2/c^4$. 
The bins with an expected 
population of less than 50 events are combined with adjacent ones. We find  
$\chi^2/ndf=2.35$ for 1065 degrees of freedom ($ndf$), which is large. 
We find that the main features of the 
Dalitz plot are well-reproduced, with some significant but numerically
small discrepancies at peaks and dips of the distribution. 
In our final results we include a conservative contribution to the 
systematic error due to uncertainties in the $\overline{D}{}^0$ decay model. 

\begin{table}
\caption{Fit results for \dkpp decay. Errors are statistical only.}
\label{dkpp_table}
\scalebox{0.94}{
\begin{tabular}{|l|c|c|c|} \hline
Intermediate state           & Amplitude 
			     & Phase ($^{\circ}$) 
                             & Fit fraction (\%)
			     \\ \hline

$K_S \sigma_1$               & $1.56\pm 0.06$
                             & $214\pm 3$
                             & $11.0\pm 0.7$\phantom{$0$}
                             \\

$K_S\rho^0$                  & $1.0$ (fixed)                                 
                             & 0 (fixed)   
                             & $21.2\pm 0.5$\phantom{$0$}
                             \\

$K_S\omega$                  & $0.0343\pm 0.0008$
                             & $112.0\pm 1.3$
                             & $0.526\pm 0.014$
                             \\

$K_S f_0(980)$               & $0.385\pm 0.006$
                             & $207.3\pm 2.3$
                             & $4.72\pm 0.05$
                             \\

$K_S \sigma_2$               & $0.20\pm 0.02$
                             & \phantom{$0$}$212\pm 12$
                             & $0.54\pm 0.10$
                             \\

$K_S f_2(1270)$              & $1.44\pm 0.04$
                             & $342.9\pm 1.7$
                             & $1.82\pm 0.05$
                             \\

$K_S f_0(1370)$              & $1.56\pm 0.12$ 
                             & $110\pm 4$
                             & $1.9\pm 0.3$
                             \\

$K_S \rho^0(1450)$           & $0.49\pm 0.08$
                             & \phantom{$00$}$64\pm 11$
                             & $0.11\pm 0.04$
                             \\

$K^*(892)^+\pi^-$            & $1.638\pm 0.010$
                             & $133.2\pm 0.4$
                             & $62.9\pm 0.8$\phantom{$0$}
                             \\ 

$K^*(892)^-\pi^+$            & $0.149\pm 0.004$
                             & $325.4\pm 1.3$
                             & $0.526\pm 0.016$ 
                             \\

$K^*(1410)^+\pi^-$	     & $0.65\pm 0.05$
			     & $120\pm 4$
                             & $0.49\pm 0.07$
			     \\

$K^*(1410)^-\pi^+$	     & $0.42\pm 0.04$
			     & $253\pm 5$
                             & $0.21\pm 0.03$ 
			     \\

$K_0^*(1430)^+\pi^-$         & $2.21\pm 0.04$
                             & $358.9\pm 1.1$
                             & $7.93\pm 0.09$
                             \\

$K_0^*(1430)^-\pi^+$         & $0.36\pm 0.03$
                             & \phantom{$0$}$87\pm 4$
                             & $0.22\pm 0.04$
                             \\

$K_2^*(1430)^+\pi^-$         & $0.89\pm 0.03$
                             & $314.8\pm 1.1$
                             & $1.40\pm 0.06$
                             \\

$K_2^*(1430)^-\pi^+$         & $0.23\pm 0.02$
                             & $275\pm 6$
                             & $0.093\pm 0.014$
                             \\

$K^*(1680)^+\pi^-$           & $0.88\pm 0.27$
                             & \phantom{$00$}$82\pm 17$
                             & $0.06\pm 0.04$
                             \\

$K^*(1680)^-\pi^+$           & $2.1\pm 0.2$
                             & $130\pm 6$
                             & $0.30\pm 0.07$
                             \\

non-resonant                 & $2.7\pm 0.3$
                             & $160\pm 5$
                             & $5.0\pm 1.0$
                             \\ 
\hline
\end{tabular}
}
\end{table}

\section{Dalitz plot analysis of \boldmath{\bddskp} decays}

\label{dalitz_analysis}

As in our previous analysis~\cite{belle_phi3_3} and in analyses 
carried out by the BaBar collaboration~\cite{babar_phi3_2, babar_phi3_3}, 
we fit the Dalitz distributions of the 
$B^+$ and $B^-$ samples separately, using Cartesian parameters 
$x_{\pm}=r_{\pm}\cos(\pm\phi_3+\delta)$ and 
$y_{\pm}=r_{\pm}\sin(\pm\phi_3+\delta)$, where the indices ``$+$" and 
``$-$" correspond to $B^+$ and $B^-$ decays, respectively. 
In this approach the amplitude ratios ($r_+$ and $r_-$) are 
not constrained to be equal for the $B^+$ and $B^-$ samples. 
Confidence intervals in $r$, $\phi_3$ and $\delta$ are then obtained 
from the $(x_{\pm},y_{\pm})$ using a frequentist technique. The advantage
of this approach is low bias and simple distributions of the fitted 
parameters, at the price of fitting in a space with higher dimensionality 
$(x_+,y_+,x_-,y_-)$ than that of the physical parameters 
$(r, \phi_3, \delta)$; see Section~\ref{section_stat}. 

Following the procedure described in Section~\ref{event_selection},
background events for the \bdk and \bdsk, \dsndpi modes 
are classified into four components:
$e^+e^-\to q\bar{q}$ (where $q=u,d,s$), 
charm, $B\overline{B}$ (except for \bddspi) 
and \bddspi\ background.
This is a refinement of the previous analysis, where
three background components were used, without separation of the 
continuum background into $(u,d,s)$ and charm.
In the case of the \bdsk mode with \dsndg, 
the $B\overline{B}$ background is divided into events 
with combinatorial $D$, and seven types of events with real 
$D$ mesons (including modes with a neutral or charged $B$ meson 
decaying to $D^{(*)}$ and a $K$, $\pi$ or $\rho$-meson).

The distributions of each of the background components are assumed to be 
factorized into products of a Dalitz plot distribution $(m^2_+, m_-^2)$, 
and distributions in $(\mbc, \de)$, and $(\thr,\fish)$. 
The shapes of these distributions are extracted from MC simulation. 
The six-dimensional PDF used for the fit is thus expressed as
\begin{equation}
  p=\sum\limits_i p_i(m_+^2, m_-^2)p_i(\mbc, \de)p_i(\thr, \fish), 
  \label{prob_dens}
\end{equation}
where the index $i$ runs over all background contributions and signal. 
The distributions $p_i(\mbc, \de)$ and $p_i(\thr, \fish)$ are 
parameterized functions. The parameterization of $p_i(\mbc, \de)$ differs
for different components: sums of two two-dimensional Gaussian 
distributions with correlations for signal and \bddskp; products of 
the empirical shape 
proposed by the ARGUS collaboration in $\mbc$~\cite{argus} 
and a linear function in $\de$ 
for $q\bar{q}$, charm and $B\overline{B}$ components. In addition, 
the parameterization for $B\overline{B}$ background includes a product 
of a Gaussian peak in $\mbc$ and a sum of exponential and 
Gaussian distributions in $\de$. 
We represent $p_i(\thr, \fish)$ as the product of two terms: 
(1) the exponential of a fourth-degree polynomial in $\thr$, and 
(2) a sum of bifurcated Gaussian distributions in $\fish$, where 
the mean and the various widths have a polynomial dependence on 
$\thr$. The function $p_i(m_+^2, m_-^2)$ is represented by Gaussian 
smoothing of the MC data.

At the first stage of the analysis (as described in 
Section~\ref{event_selection}) we determine the relative fractions 
of each background component by performing an unbinned maximum 
likelihood fit to the 
experimental data in $M_{\rm bc}$ and $\Delta E$ 
($M_{\rm bc}$, $\Delta E$, $\thr$ and $\fish$ for \bdsk, \dsndg). 
The free parameters in the fit are the fractions of 
continuum, $B\overline{B}$, and \bddspi events. The relative 
fractions of the $(u,d,s)$ and charm components of the continuum 
background, and the relative fractions of $B\overline{B}$
backgrounds with real $D^0$ for \bdsk, \dsndg mode are fixed from 
the MC simulation.

At the second stage, separate Dalitz distributions are formed 
for the $B^+$ and $B^-$ samples 
with the signal requirement for $\mbc$ and $\de$ 
($\mbc>5.27$ GeV/$c^2$, $|\de|<30$ MeV) and no requirements for 
$\thr$ and $\fish$. In each case, a fit with free parameters 
$x$ and $y$ is performed with the unbinned maximum likelihood technique,
using variables $m^2_+$, $m^2_-$, $M_{\rm bc}$, $\Delta E$, $\thr$ and 
$\fish$; only the first four variables 
were used in the previous analysis~\cite{belle_phi3_3}. 
Possible deviations from the factorization assumption 
for the background distribution and disagreements 
between MC and experimental background densities are treated in the 
systematic error. The efficiency variation as a function of the Dalitz 
plot variables is obtained from signal MC simulation and is taken into 
account in the likelihood function.

To test the consistency of the fit, the same procedure as 
used for \bddskp\ signal was applied to the \bddspip\ control 
samples. 
For the \bdpi\ and \bdspi\ (\dsndpi) modes, the results are consistent 
with the expected value $r\sim 0.01$ for the amplitude ratio. 
For \bdspi\ (\dsndg), we find $r=0.05\pm 0.02$, which is larger than 
the expected value by two standard deviations. Inspection of the Dalitz 
distributions shows visible differences between $B^+$ and $B^-$ data 
in this mode: we interpret the large value of $r$ as a statistical 
fluctuation.

\begin{table*}
  \caption{Results of the signal fits in parameters $(x,y)$. The first error 
  is statistical, and the second is experimental systematic error. 
  Statistical correlation coefficients between $x$ and $y$ are also shown.
  Model uncertainty is not included. }
  \label{sig_fit_table}
  \begin{tabular}{|l|c|c|c|}
  \hline
  Parameter  & \bdkp & \bdskp, $D^*\to D\pi^0$ & \bdskp, $D^*\to D\gamma$  \\ 
  \hline
  $x_-$ & $+0.105\pm 0.047\pm 0.011$ 
        & $+0.024\pm 0.140\pm 0.018$ 
        & $+0.144\pm 0.208\pm 0.025$ \\
  $y_-$ & $+0.177\pm 0.060\pm 0.018$ 
        & $-0.243\pm 0.137\pm 0.022$ 
        & $+0.196\pm 0.215\pm 0.037$ \\
  $x_--y_-$ correlation & $-0.289$ 
        & $+0.440$
        & $-0.207$ \\
  $x_+$ & $-0.107\pm 0.043\pm 0.011$ 
        & $+0.133\pm 0.083\pm 0.018$ 
        & $-0.006\pm 0.147\pm 0.025$ \\
  $y_+$ & $-0.067\pm 0.059\pm 0.018$ 
        & $+0.130\pm 0.120\pm 0.022$ 
        & $-0.190\pm 0.177\pm 0.037$ \\ 
  $x_+-y_+$ correlation & $+0.110$
        & $-0.101$
        & $+0.080$ \\
  \hline
  \end{tabular}
\end{table*}

\begin{figure}
  \epsfig{figure=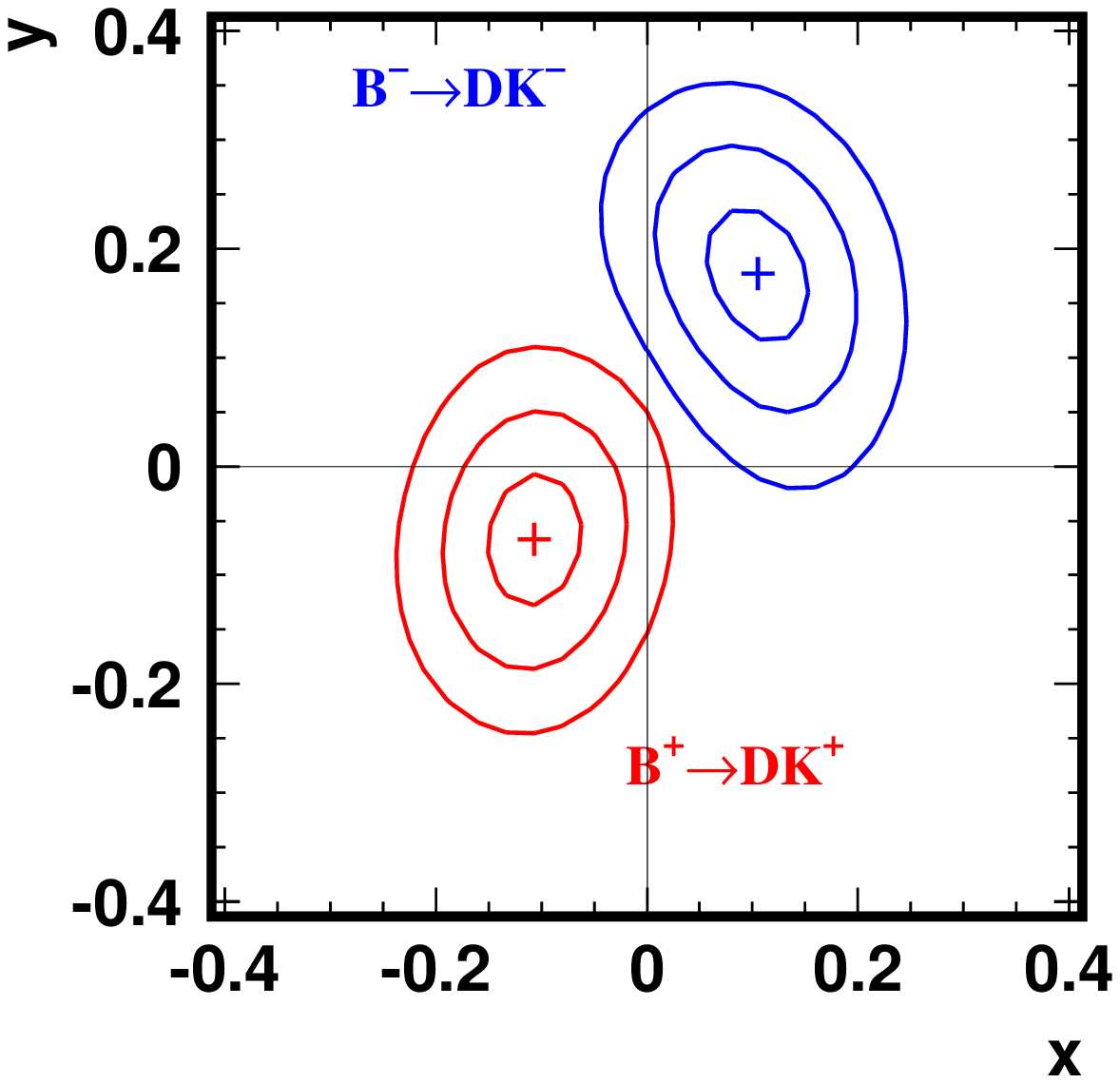,width=0.23\textwidth}
  \put(-25,75){\mbox{(a)}}
  \epsfig{figure=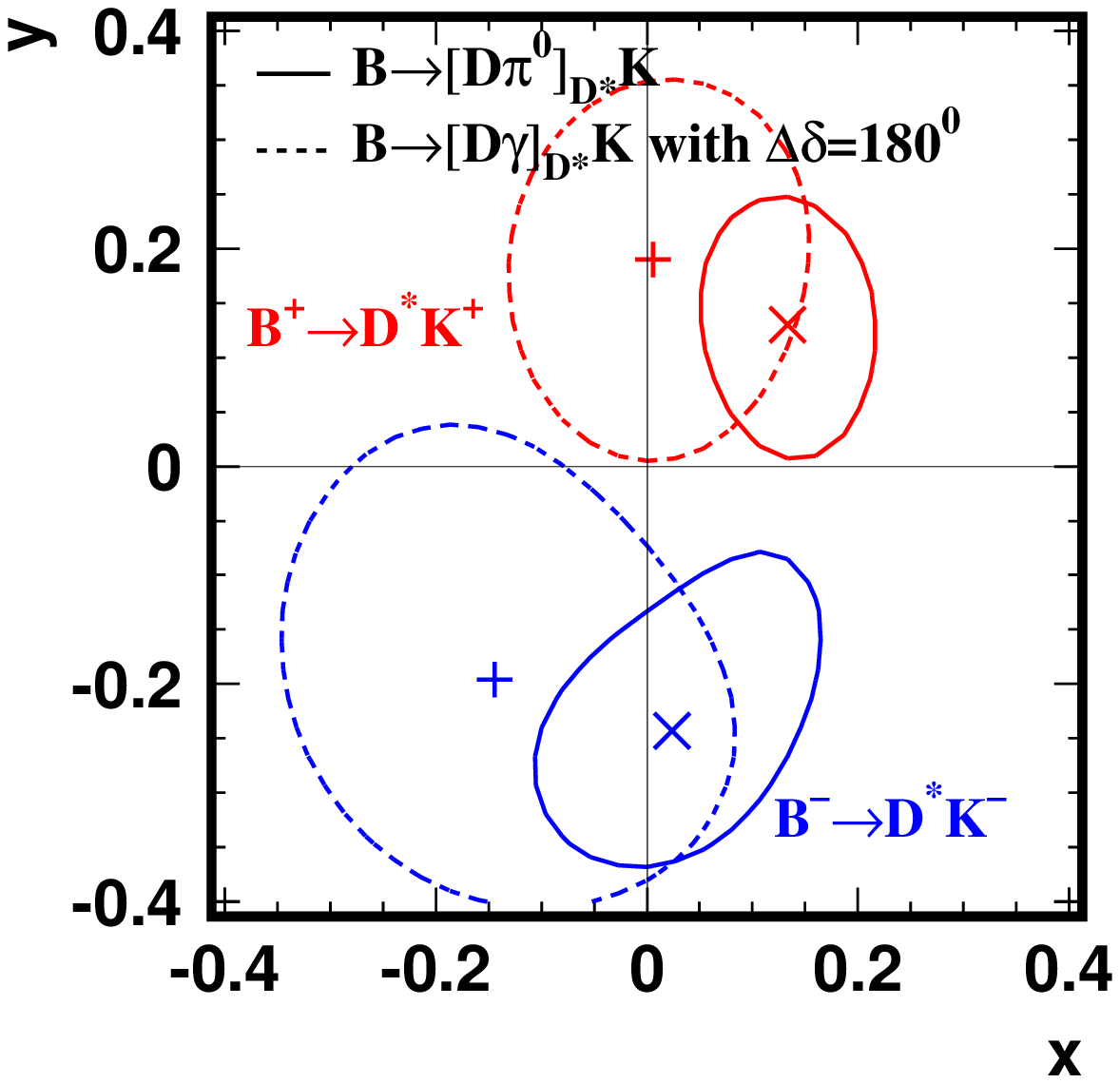,width=0.23\textwidth}
  \put(-25,75){\mbox{(b)}}

  \caption{Results of signal fits with free parameters 
           $x=r\cos\theta$ and $y=r\sin\theta$ for \bdkp\ (a), and 
           \bdskp\ (b) samples, separately for $B^-$
           and $B^+$ data. Contours indicate one, two and three
           (for \bdk) and one
           standard deviation regions (for \bdsk) obtained in the maximum 
           likelihood fit. For the \bdsk, \dsndg mode, the sign 
           of $x_{\pm}$ and $y_{\pm}$ is swapped to account for the
           relative strong phase difference of $180\deg$ with respect to
           the \bdsk, \dsndpi sample. }
  \label{sig_fit}
\end{figure}

The results of the separate $B^+$ and $B^-$ data fits are shown in 
Fig.~\ref{sig_fit}. 
The values of the fit parameters $x_{\pm}$ and $y_{\pm}$ are 
listed in Table~\ref{sig_fit_table}. As expected, the values of 
$x_{\pm}$ and $y_{\pm}$ for the \dsndg and \dsndpi modes from \bdsk 
agree within the statistical errors after reversing the signs 
in one of the modes.

\section{Evaluation of the statistical errors}

\label{section_stat}

We use a frequentist technique to evaluate the 
statistical significance of the measurements. The procedure is identical 
to that in our previous analysis~\cite{belle_phi3_3}. 
This method requires knowledge of the probability density function (PDF) of the 
reconstructed parameters $x$ and $y$ as a function of the true parameters
$\bar{x}$ and $\bar{y}$. To obtain this PDF, we use a
simplified MC simulation of the experiment which incorporates a
maximum likelihood fit with the same efficiencies, resolution 
and backgrounds as used in the fit to the experimental data. 

\begin{table*}
  \caption{$CP$ fit results. The first error is statistical, the second 
  is experimental systematic, and the third is the model uncertainty. }
  \label{fit_res_table}
  \begin{tabular}{|l||c|c|} \hline
  Parameter & \bdkp\ mode
            & \bdskp\ mode\\ \hline
  $\phi_3$  & $(80.8^{+13.1}_{-14.8}\pm 5.0\pm 8.9)^{\circ}$ 
            & $(73.9^{+18.9}_{-20.2}\pm 4.2\pm 8.9)^{\circ}$ 
            \\
  $r$       & $0.161^{+0.040}_{-0.038}\pm 0.011 ^{+0.050}_{-0.010}$
            & $0.196^{+0.073}_{-0.072}\pm 0.013 ^{+0.062}_{-0.012}$
            \\
  $\delta$  & $(137.4^{+13.0}_{-15.7}\pm 4.0\pm 22.9)^{\circ}$
            & $(341.7^{+18.6}_{-20.9}\pm 3.2\pm 22.9)^{\circ}$
            \\
  \hline
  \end{tabular}
\end{table*}

\begin{figure}
  \epsfig{figure=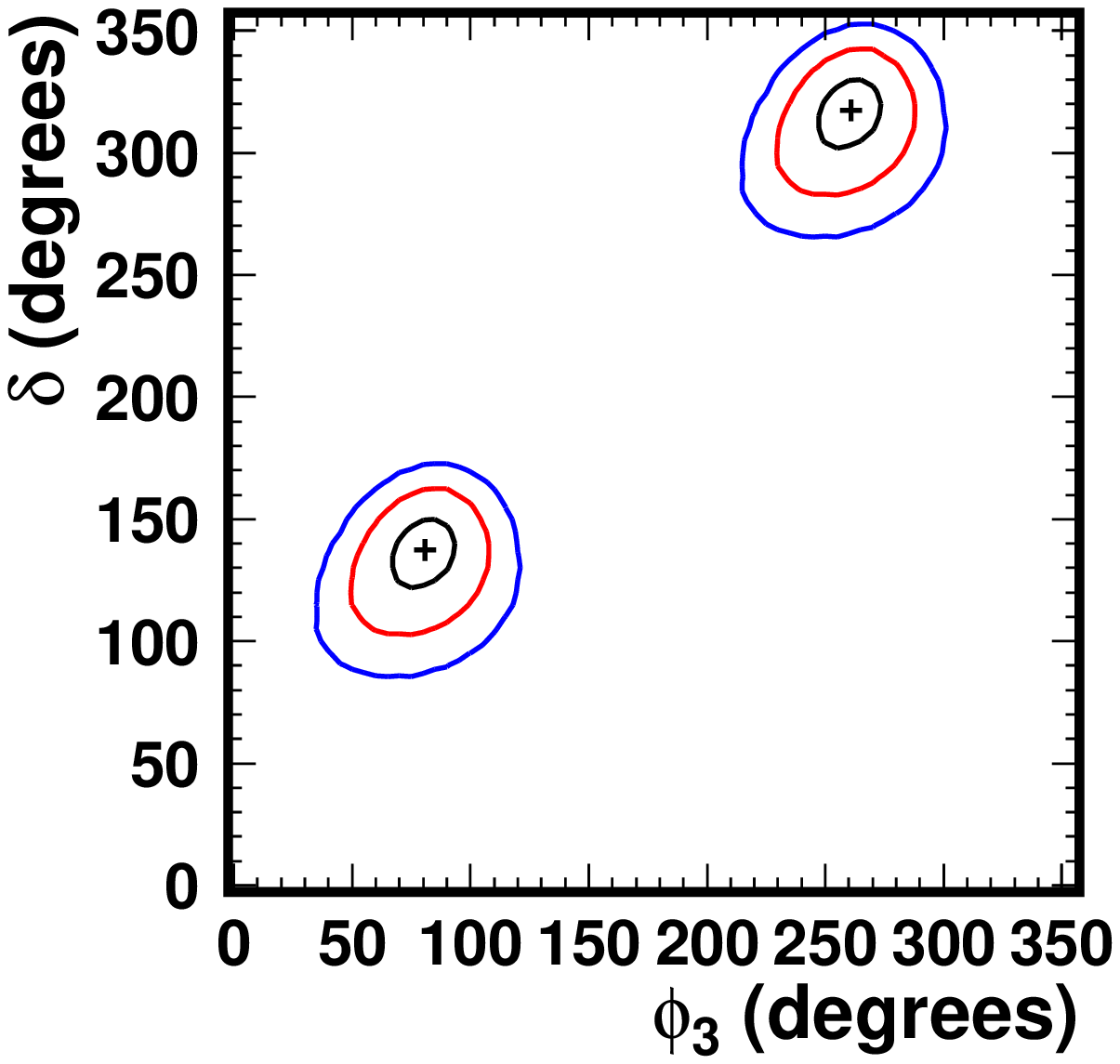,width=0.23\textwidth}
  \put(-23,95){\mbox{(a)}}
  \epsfig{figure=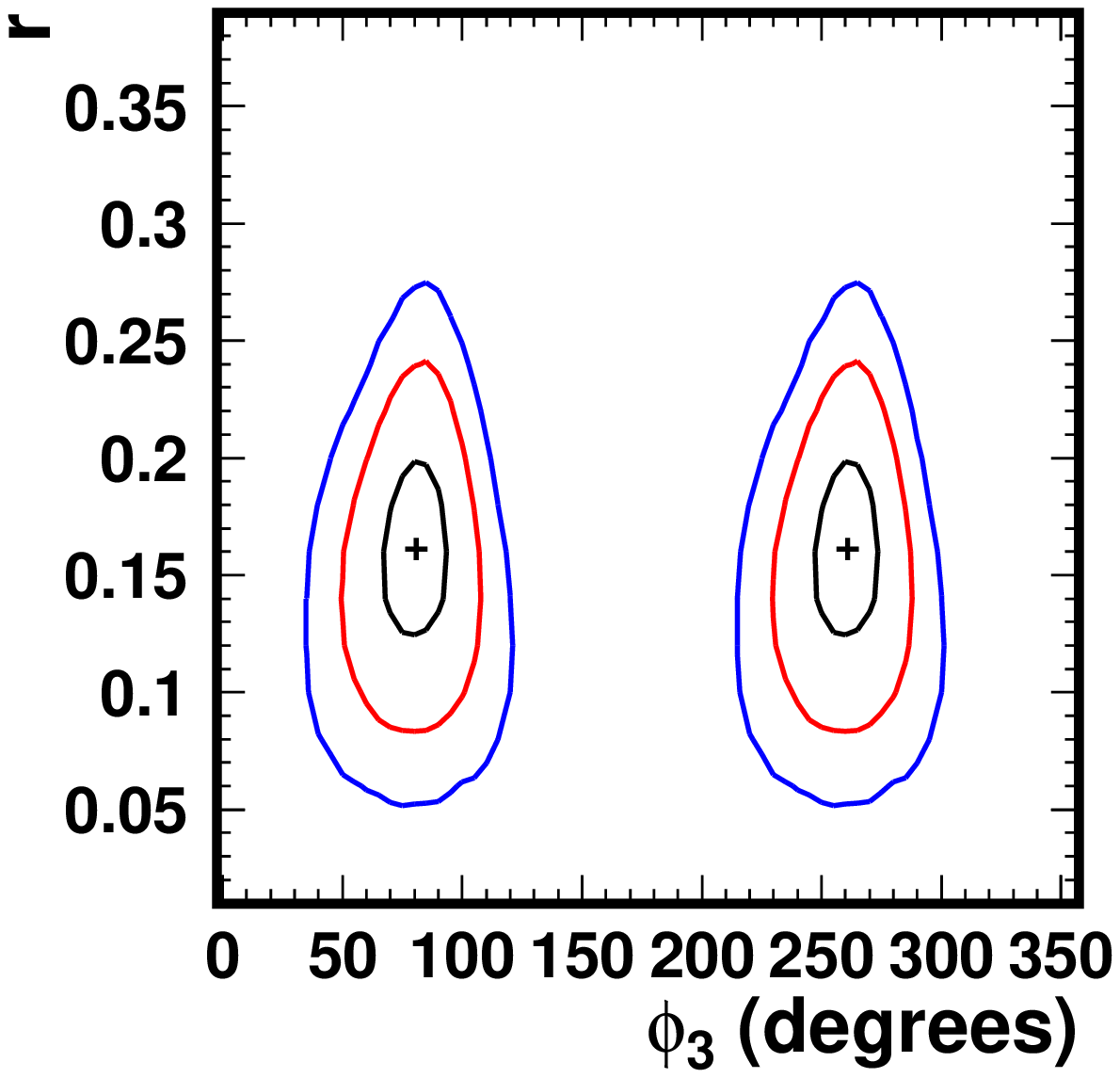,width=0.23\textwidth}
  \put(-23,95){\mbox{(b)}}

  \epsfig{figure=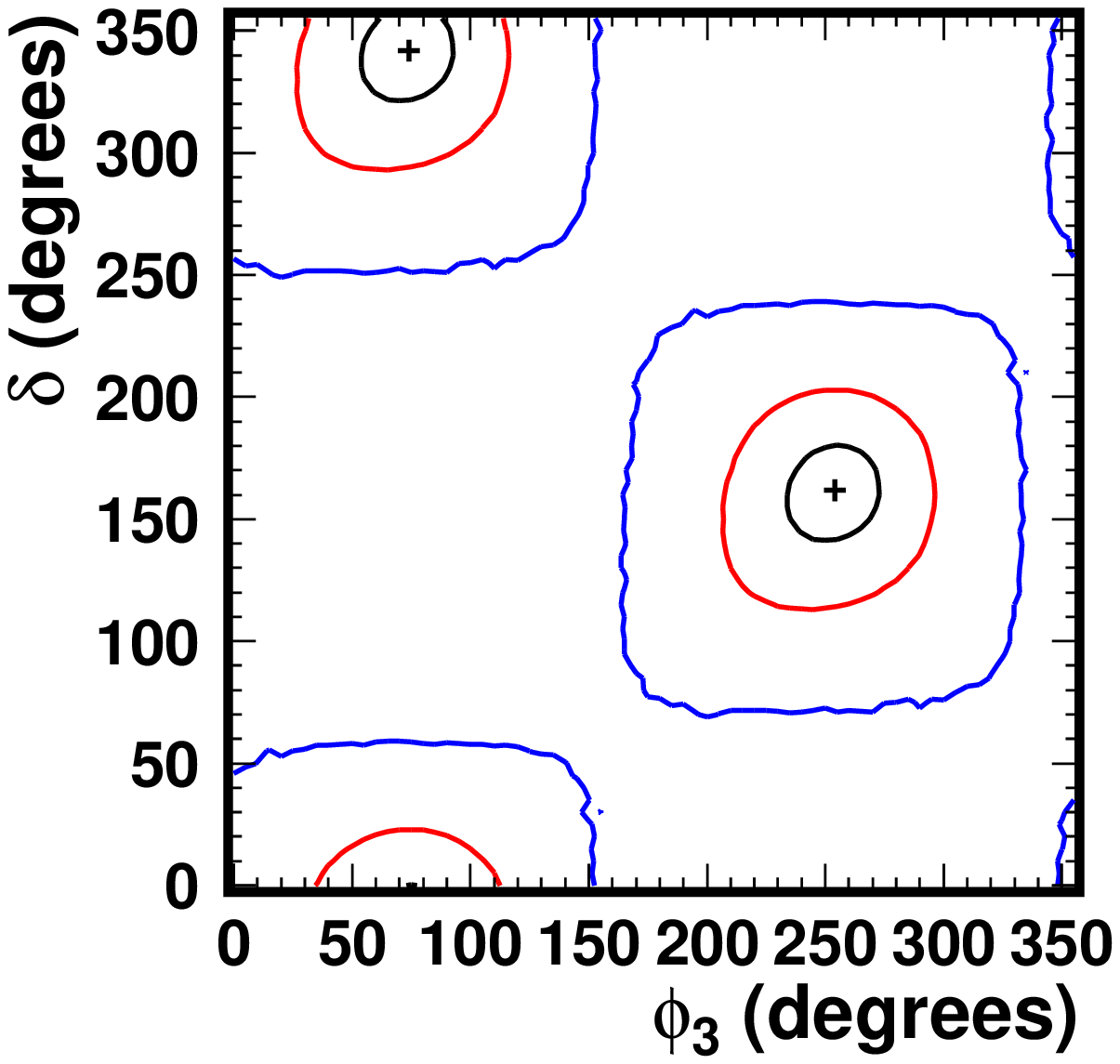,width=0.23\textwidth}
  \put(-23,95){\mbox{(c)}}
  \epsfig{figure=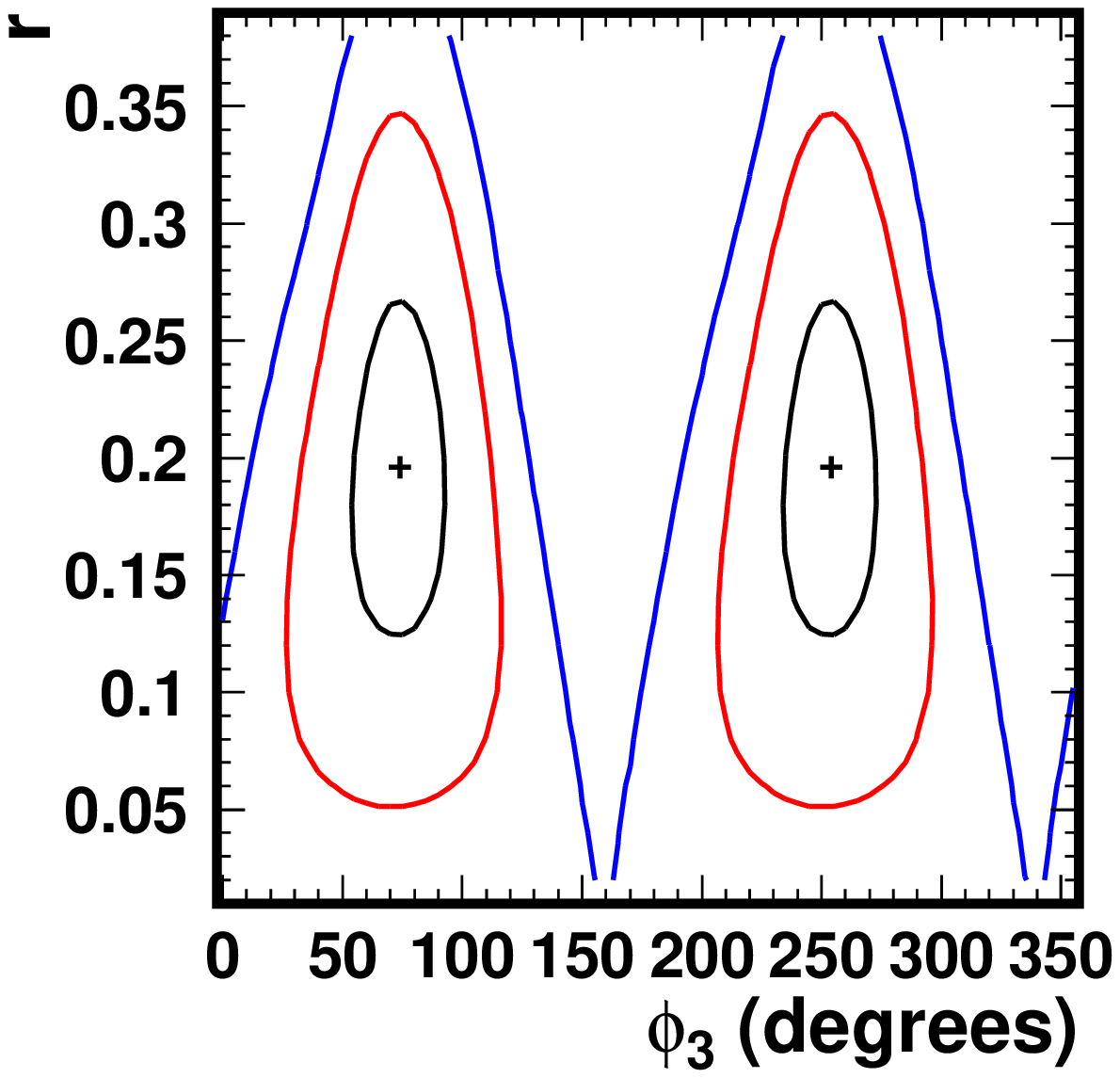,width=0.23\textwidth}
  \put(-23,95){\mbox{(d)}}

  \caption{\small Projections of confidence regions for the \bdkp (a,b) and 
           \bdskp (c,d) 
           mode onto the $(r, \phi_3)$ and $(\phi_3, \delta)$ planes. 
           Contours indicate projections of one, two and three standard 
           deviation regions. }
  \label{cont_fc}
\end{figure}

Figure~\ref{cont_fc} shows the projections of the three-dimensional 
confidence regions onto the $(r, \phi_3)$ and $(\phi_3, \delta)$ planes
for the \bdk and \bdsk modes. In the results for the \bdsk mode, we combine
both \dsndpi and \dsndg final states, taking into account the relative 
strong phase of $180\deg$ between them by swapping the sign of the
$x,y$ parameters for the \dsndg mode. 
We show the 20\%, 74\% and 97\% confidence level regions, 
which correspond to one, two, and three standard deviations for a 
three-dimensional Gaussian distribution. The values of the 
parameters $r$, $\phi_3$ and $\delta$ obtained for the \bdk and 
\bdsk modes separately are given in Table~\ref{fit_res_table}. 
The values of $\phi_3$ in these modes agree within the statistical 
errors. In general, $r$ and $\delta$ may differ: our results 
for $r$ are similar for the two modes, while the $\delta$ values are
distinct.

Note that our statistical procedure gives three-dimensional 
confidence level regions. The coverage for the set of three parameters 
$(r, \phi_3, \delta)$ is exact. One-dimensional intervals are obtained 
by projecting the three-dimensional regions onto each of the parameter axes:
exact coverage for this procedure is ensured only in the case of Gaussian
errors. In our case, Gaussian behavior of the errors is reached when 
$\sigma(r)\ll r$, and undercoverage (effectively, underestimation of 
statistical errors)
occurs if $\sigma(r)\sim r$.
The amount of undercoverage depends on the true value, $\bar{r}$: 
errors are underestimated by a factor ranging from 1.4 for $\bar{r}=0$, 
to 1.03 for $\bar{r}$ equal to the measured value.

\section{Estimation of systematic error}

\label{section_syst}

Experimental systematic errors come from uncertainty in the knowledge 
of the distributions used in the fit ({\it i.e.} Dalitz plot distributions 
of the background components, and the $(\mbc,\de)$ and 
$(\thr,\fish)$ distributions 
of the backgrounds and signal), fractions of different background components, 
and the distribution of the efficiency across the Dalitz plot. 
Uncertainties in background shapes are estimated by using alternative 
distributions in the fit (extracted from experimental data where possible). 
Uncertainties in the background fractions are obtained by varying each 
fraction within its error. Possible correlations in the 
distributions for background components that are not described by 
the formula (\ref{prob_dens}) are estimated by using independent 
background distributions in the bins of $\mbc$, $\de$, $\thr$ and 
$\fish$ variables.

In case of \bdsk, \dsndg\ decay, an additional uncertainty arises from 
the significant cross-feed from the \bdsk, \dsndpi\ mode.
The baseline \dsndg\ fit uses $x,y$ values obtained from the \bdsk, \dsndpi\ 
fit for modelling the \dsndpi\ cross-feed; to estimate
the systematic uncertainty, we vary $x,y$ within their 
errors and also take $x=y=0$. As an additional check, we apply a 
\dsndpi\ veto to the \bdsk, \dsndg\ sample: the results of this fit
are consistent with the baseline results within statistical errors.

The procedure for estimating the uncertainty due to 
the detection efficiency is different from that in the previous 
analysis~\cite{belle_phi3_3}: here we use an 
alternative efficiency shape obtained by MC simulation from the 
parameterized track finding efficiency (extracted from experimental 
data) as a function of transverse momentum and polar angle $\theta$. 

Compared to our previous analysis~\cite{belle_phi3_3}, an additional 
source of systematic error exists due to the use of $\thr$ and $\fish$
variables in the fit. However, the use of these variables increases the 
effective signal-to-background ratio, so the total systematic error
is comparable. 

Systematic errors in the physical parameters $r$, $\phi_3$
and $\delta$ are calculated from the systematic errors on the 
fitted parameters $(x,y)$. Values $(x,y)$ are generated according to 
Gaussian distributions with standard deviations equal to the 
corresponding total systematic errors; parameters $r$, $\phi_3$
and $\delta$ are then obtained for each $(x,y)$ set, and the root-mean-square 
deviations (RMS) of the resulting values are calculated. 
We perform this procedure in two ways: without correlation of 
$(x,y)$ biases for $B^+$ and $B^-$, and with 100\% correlation between them. 
The larger RMS of the two options is chosen 
as the systematic error. The systematic errors in the $x,y$ variables 
are shown in Table~\ref{sig_fit_table}. 

The model used for the \dkpp\ decay amplitude is one of the main sources of 
error for our analysis: we list this contribution separately. 
The model uncertainty splits into two contributions: one
due to imperfect description of the observable $D^0$ Dalitz plot distribution, 
and one due to uncertainty of the phase of the complex amplitude $f$, 
which is based purely on the model assumptions and appears even in the 
case of perfect description of the experimental $\overline{D}{}^0$ data.
To estimate the former contribution, we use model variations that 
give a similar \dkpp fit quality to that of the default model. 
For the latter
contribution, we take the complex phase of $f(m^2_+,m^2_-)$ from models with a
reduced number of resonances as in the previous analysis~\cite{belle_phi3_3}
while keeping the absolute value of the amplitude the same as in the 
default model. The total model uncertainty, 
$\Delta\phi_3=8.9^{\circ}$, is dominated by the
uncertainty due to complex phase. Note that
the model errors on $r$ are highly asymmetric.
While imperfect description of the $D^0$ density can lead to
a bias in both directions, a wrong complex phase introduces 
a bias only to lower values.

Our estimate of the model uncertainty can be considered
conservative. When the various $S$-wave terms --- the most
theoretically controversial part of the model --- are
replaced by a $K$-matrix amplitude~\cite{kmatrix}, the change in $\phi_3$ 
from the baseline fit does not exceed $2^{\circ}$. However, we
retain our default $8.9^{\circ}$ uncertainty as the $K$-matrix
describes only part of the amplitude.

Using a different approach, it is possible to remove the current 
model uncertainty, exploiting constraints on the complex phase 
in the \dkpp amplitude that can be obtained 
experimentally from the analysis of $\psi(3770)\to D^0\overline{D}{}^0$ decays.  
Such a measurement was recently performed by CLEO~\cite{cleo_modind}. 
The results show good agreement with the isobar model, however a
quantitative estimate of the model uncertainty for a model-dependent 
fit is hard to obtain from these data. Instead, a model-independent 
analysis~\cite{giri,modind,modind2} involving a binned fit of the \dkpp Dalitz distribution
is possible. The model error in this analysis will be replaced by a statistical 
error of about 1--2$^{\circ}$ due to the finite $\psi(3770)\to D^0\overline{D}{}^0$ 
sample, while the statistical error 
associated with the $B$ data sample should increase by 10-20\% due to
the binned fit procedure. At the current level of precision, this will not 
result in a significant improvement in the precision of $\phi_3$, 
but future analyses with larger samples of $B$ decays should benefit from 
the model-independent technique.

\section{Combined \boldmath{$\phi_3$} measurement}

\label{section_combined}

The two event samples, \bdkp\ and \bdskp, are combined 
in order to improve the sensitivity to $\phi_3$. 
The confidence levels for the combination of the two modes are obtained 
using the same frequentist technique as for the 
single mode, with the PDF of the two measurements being the product 
of the probability densities for the individual modes. 
Confidence intervals for the combined measurement together 
with systematic and model errors are shown in Table~\ref{fc_comb_table}. 
The statistical confidence level of $CP$ violation 
is $1-CL=1.5\times 10^{-4}$, or 3.8 standard deviations.
With the systematic and model errors taken into account, 
$CP$ conservation is ruled out at the 
confidence level $5\times 10^{-4}$, or 3.5 standard deviations.
The systematic errors are assumed to 
be uncorrelated in this calculation; the resulting estimate is 
conservative, as
most of the systematic biases are correlated between $B^+$ and $B^-$ samples
and thus do not introduce $CP$ violation.

\begin{table*}
  \caption{Results of the combination of the \bdkp\ and \bdskp\ modes. }
  \label{fc_comb_table}
  \begin{tabular}{|l|c|c|c|c|c|} \hline
  Parameter   & $1\sigma$ interval & $2\sigma$ interval & 
               Systematic error & Model uncertainty \\ \hline
  $\phi_3$    & $(78.4^{+10.8}_{-11.6})\deg$ 
              & $54.2\deg<\phi_3<100.5\deg{}$
              & $3.6^{\circ}$ & $8.9^{\circ}$ \\
  $r_{DK}$    & $0.160^{+0.040}_{-0.038}$
              & $0.084<r_{DK}<0.239$
              & $0.011$ & $^{+0.050}_{-0.010}$ \\
  $r_{D^*K}$  & $0.196^{+0.072}_{-0.069}$ 
              & $0.061 < r_{D^*K} < 0.271$
              & $0.012$ & $^{+0.062}_{-0.012}$ \\
  $\delta_{DK}$   & $(136.7^{+13.0}_{-15.8})\deg$ 
                  & $102.2\deg < \delta_{DK} < 162.3\deg$
                  & $4.0^{\circ}$ & $22.9^{\circ}$ \\
  $\delta_{D^*K}$ & $(341.9^{+18.0}_{-19.6})\deg$ 
                  & $296.5\deg < \delta_{D^*K} < 382.7\deg$
                  & $3.0^{\circ}$ & $22.9^{\circ}$ \\
  \hline
  \end{tabular}
\end{table*}

\section{Conclusion}

We report the results of a measurement of the unitarity triangle angle 
$\phi_3$, using a method based on Dalitz plot analysis of
\dkpp\ decay in the process \bddskp. A new measurement of 
$\phi_3$ using this technique was performed based on 605 fb$^{-1}$ 
of data collected by the Belle detector: 70\% larger than the previous 
sample~\cite{belle_phi3_3}. The statistical sensitivity of the measurement has also been 
improved by modifications to the event selection and fit procedure, and 
by adding the sample with $D^*$ decaying to the $D\gamma$ final state. 

From the combination of \bdkp and \bdskp modes, we obtain the value
$\phi_3=78.4\deg{}^{+10.8\deg}_{-11.6\deg}
\mbox{(stat)}\pm 3.6\deg \mbox{(syst)}\pm 8.9\deg(\mbox{model})$; 
of two possible solutions we have chosen the one with $0<\phi_3<180^{\circ}$.
We also obtain values of the amplitude ratios 
$r_{DK}=0.160^{+0.040}_{-0.038}\mbox{(stat)}\pm
0.011\mbox{(syst)}^{+0.050}_{-0.010}\mbox{(model)}$, and 
$r_{D^*K}=0.196^{+0.072}_{-0.069}\mbox{(stat)}\pm 0.012\mbox{(syst)}^{+0.062}_{-0.012}\mbox{(model)}$. 
The $CP$ conservation in the combined measurement is ruled out at the 
confidence level $5\times 10^{-4}$, or 3.5 standard deviations.

The statistical precision of the $\phi_3$ measurement is already
comparable to the estimated model uncertainty. However, it is possible
to eliminate this model uncertainty using constraints on the
\dkpp decay amplitude obtained by the CLEO collaboration in the 
analysis of $\psi(3770)\to
D^0\overline{D}{}^0$ decays~\cite{modind,modind2,cleo_modind}.
The statistical errors in the proposed binned fit procedure 
are 10-20\% larger, but the model uncertainty is replaced 
by a small ($1-2^{\circ}$) statistical error due to the finite 
$\psi(3770)\to D^0\overline{D}{}^0$ sample. This should 
result in an improvement of the $\phi_3$ precision in 
future high-statistics analyses.

\section*{Acknowledgments}

We thank the KEKB group for the excellent operation of the
accelerator, the KEK cryogenics group for the efficient
operation of the solenoid, and the KEK computer group and
the National Institute of Informatics for valuable computing
and SINET3 network support.  We acknowledge support from
the Ministry of Education, Culture, Sports, Science, and
Technology (MEXT) of Japan, the Japan Society for the 
Promotion of Science (JSPS), and the Tau-Lepton Physics 
Research Center of Nagoya University; 
the Australian Research Council and the Australian 
Department of Industry, Innovation, Science and Research;
the National Natural Science Foundation of China under
contract No.~10575109, 10775142, 10875115 and 10825524; 
the Ministry of Education, Youth and Sports of the Czech 
Republic under contract No.~LA10033;
the Department of Science and Technology of India; 
the BK21 and WCU program of the Ministry Education Science and
Technology, National Research Foundation of Korea,
and NSDC of the Korea Institute of Science and Technology Information;
the Polish Ministry of Science and Higher Education;
the Ministry of Education and Science of the Russian
Federation and the Russian Federal Agency for Atomic Energy;
the Slovenian Research Agency;  the Swiss
National Science Foundation; the National Science Council
and the Ministry of Education of Taiwan; and the U.S.\
Department of Energy.
This work is supported by a Grant-in-Aid from MEXT for 
Science Research in a Priority Area (``New Development of 
Flavor Physics"), and from JSPS for Creative Scientific 
Research (``Evolution of Tau-lepton Physics").


\begin{thebibliography}{100}

\bibitem{ckm}
M. Kobayashi and T. Maskawa, Prog. Theor. Phys. {\bf 49}, 652 (1973); 
N. Cabibbo, Phys. Rev. Lett. {\bf 10}, 531 (1963). 

\bibitem{glw}
M. Gronau and D. London, Phys. Lett. {\bf B253}, 483 (1991);
M. Gronau and D. Wyler, Phys. Lett. {\bf B265}, 172 (1991).

\bibitem{dunietz}
I. Dunietz, Phys. Lett. {\bf B270}, 75 (1991).

\bibitem{eilam}
D. Atwood, G. Eilam, M. Gronau and A. Soni, Phys. Lett. {\bf B341}, 372 (1995).

\bibitem{ads}
D. Atwood, I. Dunietz and A. Soni, Phys. Rev. Lett. {\bf 78}, 3257 (1997);
D. Atwood, I. Dunietz and A. Soni, Phys. Rev. D {\bf 63}, 036005 (2001).

\bibitem{bigi}
I. I. Bigi and A. I. Sanda, Phys. Lett. {\bf B211}, 213 (1988);
A. B. Carter and A. I. Sanda, Phys. Rev. Lett {\bf 45}, 952 (1980). 

\bibitem{giri}
A. Giri, Yu. Grossman, A. Soffer, J. Zupan, Phys. Rev. D {\bf 68},
054018 (2003).

\bibitem{binp_dalitz}
A. Bondar. Proceedings of BINP Special Analysis Meeting on Dalitz Analysis, 
24-26 Sep. 2002, unpublished. 

\bibitem{wolfenstein}   
L.~Wolfenstein,
Phys. Rev. Lett. {\bf 51}, 1945 (1983).

\bibitem{mixing_hfag} 
A.~J.~Schwartz (for HFAG charm group),
arXiv:0911.1464 [hep-ex].
                                                                                
\bibitem{mixing_phi3} Y. Grossman, A. Soffer, J. Zupan, Phys. Rev. D {\bf 72},  
031501 (2005).

\bibitem{belle_phi3_3}
Belle Collaboration, A.~Poluektov {\it et al.}, Phys. Rev. D {\bf 73}, 112009 (2006).

\bibitem{belle_phi3_1}
Belle Collaboration, K.~Abe {\it et al.}, arXiv:hep-ex/0308043. 

\bibitem{belle_phi3_2}
Belle Collaboration, A.~Poluektov {\it et al.}, Phys. Rev. D {\bf 70}, 072003 (2004).

\bibitem{babar_phi3_2}
BaBar Collaboration, B.~Aubert {\it et al.}, Phys. Rev. Lett. {\bf 95}, 
121802 (2005).

\bibitem{babar_phi3_3}
BaBar Collaboration, B.~Aubert {\it et al.}, Phys. Rev. D {\bf 78}, 
034023 (2008). 

\bibitem{babar_3pi}
BaBar Collaboration, B. Aubert {\it et al.}, Phys. Rev. Lett. {\bf 99}, 251801 (2007). 

\bibitem{belle}
Belle Collaboration, A.~Abashian {\it et al.}, Nucl. Instr. and Meth. A {\bf 479}, 117 (2002).

\bibitem{svd2}
Y. Ushiroda (Belle SVD2 Group), Nucl. Instr. and Meth. A {\bf 511}, 6 (2003). 

\bibitem{bondar_gershon}
A. Bondar and T. Gershon, Phys. Rev. D {\bf 70}, 091503 (2004). 

\bibitem{fisher}
CLEO Collaboration, D. M. Asner {\it et al.}, Phys. Rev. D {\bf 53}, 1039
(1996).

\bibitem{pdg}
Particle Data Group, C. Amsler {\it et al.}, Phys. Lett. B {\bf 667}, 1 (2008).

\bibitem{kmatrix}
V. V. Anisovich and A. V. Sarantsev, Eur. Phys. J. A {\bf 16}, 229 (2003). 

\bibitem{cleo_model}
CLEO Collaboration, S. Kopp {\it et al.}, Phys. Rev. D {\bf 63}, 092001 (2001).

\bibitem{dkpp_cleo}
CLEO Collaboration, H. Muramatsu {\it et al.}, Phys. Rev. Lett.
{\bf 89}, 251802 (2002), Erratum-ibid: {\bf 90}, 059901 (2003).

\bibitem{aitala2}
E791 Collaboration, E. M. Aitala {\it et al.}, Phys. Rev. Lett. {\bf 86}, 770 (2001).

\bibitem{argus}
ARGUS Collaboration, H.~Albrecht {\it et al.}, Phys. Lett. B {\bf 241}, 278 (1990).

\bibitem{cleo_modind} 
CLEO Collaboration, R.A. Briere {\it et al.}, 
Phys. Rev. D {\bf 80}, 032002 (2009). 

\bibitem{modind} A. Bondar, A. Poluektov, 
Eur. Phys. J. C {\bf 47}, 347 (2006). 

\bibitem{modind2}
A. Bondar, A. Poluektov, Eur. Phys. J. C {\bf 55}, 51 (2008).

\end{thebibliography}
\end{document}